\documentclass[%
reprint,
amsmath,amssymb,
aps,
]{revtex4-2}

\usepackage{amsthm}
\usepackage{graphicx}
\usepackage{dcolumn}
\usepackage{bm}
\usepackage{tikz}
\usetikzlibrary{quantikz}
\usepackage{physics}
\usepackage{adjustbox}
\usepackage{subfigure}
\usepackage{pgfplots}
\usepackage[ruled,lined,linesnumbered]{algorithm2e}
\usepackage{multirow}
\usepackage{hyperref}
\usepackage{threeparttable}
\usepackage{enumitem}
\newtheorem{definition}{Definition}
\newtheorem{lemma}{Lemma}

\theoremstyle{definition}  
\newtheorem{example}{Example}
\definecolor{mycolor1}{RGB}{057,081,162}
\definecolor{mycolor2}{RGB}{092,144,194}
\definecolor{mycolor3}{RGB}{253,185,107}
\definecolor{mycolor4}{RGB}{218,056,042}
\definecolor{mycolor5}{RGB}{212,212,212}
\definecolor{deepgreen}{RGB}{0,100,0}

\begin{document}
	
	\preprint{APS/123-QED}
	
	\title{Nearest neighbor synthesis of CNOT circuits on general quantum architectures}

	\author{Xinyu Chen}
	\thanks{Xinyu Chen and Mingqiang Zhu contributed equally to this work.}
	\affiliation{
		School of Information Science and Technology, Nantong University, Nantong 226019, China
	}%
	\author{Mingqiang Zhu}
    \thanks{Xinyu Chen and Mingqiang Zhu contributed equally to this work.}
	\affiliation{
		School of Information Science and Technology, Nantong University, Nantong 226019, China
	}%
    
	\author{Xueyun Cheng}
	\email{chen.xy@ntu.edu.cn}
	\affiliation{
		School of Information Science and Technology, Nantong University, Nantong 226019, China
	}%
     \author{Shiguang Feng}
    \affiliation{%
     School of Computer Science and Engineering, Sun Yat-sen University, Guangzhou 510006, China
    }%
     \author{Pengcheng Zhu}
    \affiliation{%
    	Department of Artificial Intelligence, Suqian University, Suqian, 223800, China
    }%
	\author{Zhijin Guan}
    \affiliation{
       School of Information Science and Technology, Nantong University, Nantong 226019, China
    }%

	\date{\today}
	
	\begin{abstract}
        NISQ devices have inherent limitations in terms of connectivity and hardware noise. The synthesis of CNOT circuits considers the physical constraints and transforms quantum algorithms into low-level quantum circuits that can execute on physical chips correctly. In the current trend, quantum chip architectures without Hamiltonian paths are gradually replacing architectures with Hamiltonian paths due to their scalability and low-noise characteristics. To this end, this paper addresses the nearest neighbor synthesis of CNOT circuits in the architectures with and without Hamiltonian paths, aiming to enhance the fidelity of the circuits after execution. Firstly, a key-qubit priority mapping model for general quantum architectures is proposed. Secondly, the initial mapping is further improved by using tabu search to reduce the number of CNOT gates after circuit synthesis and enhance its fidelity. Finally, the noise-aware CNOT circuit nearest neighbor synthesis algorithm for the general architecture is proposed based on the key-qubit priority mapping model. The algorithm is demonstrated on several popular cloud quantum computing platforms and simulators, showing that it effectively optimizes the fidelity of CNOT circuits compared with mainstream methods. Moreover, the method can be extended to more general circuits, thereby improving the overall performance of quantum computing on NISQ devices.

	\end{abstract}
	
	\maketitle
	

	\section{Introduction}
		Quantum computing is a new computing paradigm that follows the laws of quantum mechanics to perform complex tasks. It can provide up to exponential speedups over classical algorithms in integer factorization~\cite{peng2008quantum,jiang2018quantum}, database search~\cite{giri2017review}, quantum many-body simulation~\cite{somaroo1999quantum}, \textit{etc}. Moreover, quantum computing has potential applications in cryptography~\cite{kumar2021state}, chemistry~\cite{werner2012molpro,mcardle2020quantum}, artificial intelligence~\cite{dunjko2018machine,situ2020quantum} and other areas. In recent years, quantum computing has entered the era of Noisy Intermediate-Scale Quantum (NISQ), which supports quantum computing with tens to hundreds of qubits. However, NISQ devices are limited in their ability to apply only a few elementary quantum gates due to the presence of noise. Among these gates, the CNOT gate is widely employed in quantum computing as it can be combined with other single-qubit gates to construct a universal gate library~\cite{barenco1995elementary}. The circuits composed of CNOT gates cannot be executed directly since the nearest neighbor (NN) constraint and error rate of NISQ devices are affected by the physical architecture. Instead, these circuits must be further transformed into a suitable form that can be implemented on actual quantum hardware.

	    Currently, there are two main methods to obtain a quantum circuit that satisfies the NN constraint. The first method involves inserting SWAP gates in front of quantum gates that do not satisfy the NN constraint after the initial mapping~\cite{zhu2020exact,cheng2020nearest,zhu2023variation}. This ensures the connection of non-nearest neighbor gates on physical qubits. Since a SWAP gate can be decomposed into three CNOT gates, this method necessitates the insertion of additional CNOT gates. Consequently, the depth of the final circuit increases, raising the probability of errors in the circuit. The other approaches focus on researching the synthesis of quantum circuits directly under NN constraints. Quantum circuit synthesis enables the transformation of a matrix representing a quantum algorithm into a quantum circuit supporting a specific library of gates~\cite{nash2020quantum,amy2018controlled,zhang2023characterization,cheng2022nearest,zhang2022automatic,gheorghiu2022reducing}. The synthesis method that generates a circuit containing only CNOT gates is CNOT circuit synthesis. The resulting CNOT circuits do not necessarily satisfy the NN constraint of the physical architecture and cannot be executed on real quantum computing devices. To address this issue, the CNOT circuit NN synthesis method that transforms the Boolean matrix into a CNOT circuit satisfying the NN constraint of the physical architecture was proposed to solve this problem~\cite{markov2008optimal,schaeffer2012linear,de2021gaussian,kissinger2020cnot,de2020quantum,meijer2022dynamic,zhu2022physical,chen2022recursive,wu2023optimization}.
	    
	    Several researchers have dedicated their efforts to the synthesis of quantum circuits. In~\cite{markov2008optimal}, CNOT circuits were represented as Boolean matrices, and a CNOT circuit synthesis algorithm based on Gaussian elimination and LU decomposition was proposed. However, the NN constraints between control and target qubits of CNOT gates were not considered. A linear NN Gaussian elimination method was proposed in~\cite{schaeffer2012linear}, which shared a similar structure to the Gaussian elimination method. But the row operations must be performed between adjacent rows, and this method is only applicable to the NN synthesis problem of CNOT circuits in a one-dimensional (1D) structure. In~\cite{de2021gaussian} and~\cite{kissinger2020cnot}, the NN synthesis methods for CNOT circuits in a two-dimensional (2D) structure were proposed, relying on Steiner trees to determine the NN interaction paths on a 2D grid. A similar strategy was proposed in~\cite{de2020quantum}. These methods all use Steiner trees to determine the interaction paths that satisfy the NN constraint. In contrast to the traditional Gaussian elimination, they do not completely rely on the main diagonal elements in the elimination process, but they use the neighboring elements with value 1 in the Steiner tree to achieve the row elimination. In~\cite{meijer2022dynamic}, a new algorithm for CNOT circuit synthesis based on corrected sub-decoding was proposed to solve the problem of synthesizing CNOT circuits with fully connected topologies as well as quantum devices with topologically constrained structures. The authors in~\cite{zhu2022physical} aimed to reduce the effect of noise on CNOT circuit synthesis by reducing the number of CNOT gates and circuit depth, while minimizing the errors generated during the synthesis process.
	    
	    Some of the above studies focus on CNOT circuit synthesis on quantum architectures with Hamiltonian paths. A Hamiltonian path in a graph is defined as a path that visits each vertex in the graph exactly once, showcasing a potential route for optimal quantum gates arrangement. However, with the development of NISQ devices, architectures without Hamiltonian paths are gradually replacing those with Hamiltonian paths because of their excellent scalability and high fault tolerance. Consequently, the CNOT circuit synthesis methods designed for architectures with Hamiltonian paths cannot be directly applied to architectures without Hamiltonian paths. Various solutions have been proposed to address this challenge. Although the core method in \cite{kissinger2020cnot} originally targeted architectures with Hamiltonian paths, its recursive extension explicitly addresses this limitation by enabling synthesis on architectures without Hamiltonian paths.  Meanwhile, both~\cite{chen2022recursive} and~\cite{wu2023optimization} presented ROWCOL methods for implementing NN synthesis of CNOT circuits on arbitrary architectures. These methods prioritize the elimination of certain qubits and subsequently disconnect them. This allows finding suboptimal Hamiltonian paths in architectures where no Hamiltonian paths exist. Through this process, the size of the problem is gradually reduced. Another approach is to use matrix decomposition techniques~\cite{li2013decomposition} to represent the quantum algorithm as a quantum circuit. The circuit is then mapped directly onto the physical architecture, and the NN of the qubits is implemented using traditional quantum routing methods~\cite{niu2020hardware}. While this approach avoids the reliance on Hamiltonian paths, it requires more resources and time to implement the NN of quantum circuits, and may not lead to optimal solutions. In practical applications, these drawbacks may lead to lower efficiency in quantum computing or even failure to achieve the desired computational tasks.
	    
	    With enhancing the fidelity as the goal, we propose a solution for the NN synthesis of CNOT circuits on general quantum architectures. The main contributions of this paper are as follows.
	    
	    \begin{itemize}
	    	\item[$\bullet$] The challenges associated with CNOT quantum circuit synthesis on quantum architectures without Hamiltonian paths are analyzed, and a key-qubit priority mapping model is proposed to solve the problem. Based on this model, the initial mapping of key qubits is further improved by using tabu search to enhance the circuit fidelity after synthesis.
	    	\item[$\bullet$]Based on the initial mapping strategy of key qubits, an NN synthesis of CNOT circuits method is proposed for the general architecture. This method converges the Boolean matrix by layers under the premise of satisfying the NN constraint, and gradually realizes the matrix transformation, thus transforming into a CNOT circuit with a lower error rate.
	    	\item[$\bullet$]The above algorithm is implemented on several mainstream cloud quantum computing platforms, including real quantum computers and quantum simulators. The demonstration results show that, by employing noise-aware synthesis methods and reducing the number of CNOT gates, the fidelity of CNOT quantum circuits is significantly improved. Compared to current state-of-the-art synthesis methods, the fidelity is improved by 21.3\% and 7.0\% on IBMQ\_QX5 and IBMQ\_Tokyo, respectively. Compared to current mainstream cloud platform quantum compilers, the fidelity is improved by a factor of 5.7 and 243 on IBMQ\_Qutio and IBMQ\_Guadalupe, respectively. The fidelity of the Bernstein-Vazirani (BV) algorithm is improved by 14.8\% on both the quantum computing devices and simulators.
	    \end{itemize}
	    
	    This paper is organized as follows: in Section~\ref{sec:background}, basic concepts involving CNOT circuit synthesis are briefly introduced. In Section~\ref{sec:keyqubit}, key-qubit priority mapping for general quantum architectures is studied, a key-qubit priority model is established, and the key-qubit priority initial mapping is optimized by tabu search. In Section~\ref{sec:lcnns}, a noise-aware CNOT circuit NN synthesis based on the key-qubit priority model is considered, and a CNOT circuit NN synthesis algorithm applicable to the general architecture is given. In Section~\ref{sec:experimental}, the proposed method's optimization for circuit fidelity is verified by the execution results on quantum computing devices. The paper is discussed and summarized in Section~\ref{sec:discussion} and~\ref{sec:conclusion}.

	   \section{Background}
	    \label{sec:background}
	    \subsection{CNOT gate and CNOT circuit}
	    The CNOT (Controlled-NOT) gate is a basic two-qubit quantum logic gate that performs a conditional NOT operation on the target qubit depending on the state of the control qubit. Fig.~\ref{fig1} shows a graphical representation of a CNOT gate with two input qubits and its unitary matrix, where $q_0$ denotes the control bit and $q_1$ denotes the target bit. The operation of CNOT gate can be represented by a unitary matrix, which defines the relationship between input and output.
	    \begin{figure}
	    	\centering
	    	\begin{adjustbox}{width=0.48\textwidth}
	    		\begin{quantikz}[row sep={0.5cm,between origins},column sep=0.3cm]
	    			\lstick{\scriptsize $q_0$}  & \ctrl{1}     &  \qw \rstick[wires=2]{\hspace{2mm}$U_{\text{CNOT}} = \left[\begin{smallmatrix} 1 & 0 & 0 & 0 \\ 0 & 1 & 0 & 0 \\ 0 & 0 & 0 & 1 \\ 0 & 0 & 1 & 0 \end{smallmatrix}\right]$}\\
	    			\lstick{\scriptsize $q_1$}  & \targ{}    & \qw	
	    		\end{quantikz}
	    	\end{adjustbox}
	    	\caption{The CNOT gate and its unitary matrix.}
	    	\label{fig1}
	    \end{figure}

	    This unitary matrix shows that the CNOT gate flips the state of the target qubit if and only if the control qubit is in the $\ket{1}$ state, which can be expressed in Eq.~\eqref{eq2}.
	    \begin{equation}
	    	|00\rangle \rightarrow|00\rangle ;|01\rangle \rightarrow|01\rangle ;|10\rangle \rightarrow|11\rangle ;|11\rangle \rightarrow|10\rangle
	    	\label{eq2}
	    \end{equation}
	    
	    Due to the capability of representing any multi-qubit gate by using a combination of CNOT gates and single-qubit gates, CNOT gates are frequently employed in quantum circuits for qubit state manipulation. A circuit consisting only of CNOT gates is called a CNOT circuit.

	    \subsection{Quantum topology architecture}
	    Although the number of qubits in NISQ devices has increased, they still face various physical constraints, including connectivity constraints and gate errors. The connectivity between different physical qubits can be represented by a quantum topology graph. Fig.~\ref{fig2} displays the coupling graph of two IBM NISQ device topologies. Coupling graphs for all the other architectures considered in the paper are shown in the Appendix~\ref{Coupling}. Currently, the only two-qubit gate supported by IBM devices are CNOT gates, which can only operate on two qubits connected by an edge in the coupling graph. In other words, the logical qubits of the CNOT gate mapped onto these two connected physical qubits satisfy the NN constraint. Note that there is no Hamiltonian path in the two coupling graphs in Fig.~\ref{fig2}.
	    
	    \begin{figure}
	    	\centering
	    	\subfigure[]{
	    		\begin{adjustbox}{width=0.25\textwidth}
	    			\begin{tikzpicture}[>=stealth]
	    				\node at (-6,3) [circle,draw=black!100,inner sep=2.2mm] {\Large4};
	    				\node at (-6,6) [circle,draw=black!100,inner sep=2.2mm] {\Large3};
	    				\node at (-6,9) [circle,draw=black!100,inner sep=2.2mm] {\Large1};
	    				\node at (-9,9) [circle,draw=black!100,inner sep=2.2mm] {\Large0};
	    				\node at (-3,9) [circle,draw=black!100,inner sep=2.2mm] {\Large2};
	    				\draw [thick] (-3.5,9) -- (-5.5,9);
	    				\draw [thick] (-6.5,9) -- (-8.5,9);
	    				\draw [thick] (-6,3.5) -- (-6,5.5);
	    				\draw [thick] (-6,6.5) -- (-6,8.5);
	    				\draw (-5.3,4.5) node {8.791e-3};
	    				\draw (-5.3,7.5) node {7.440e-3};
	    				\draw (-4.5,9.25) node {7.768e-3};
	    				\draw (-7.5,9.25) node {1.631e-2};
	    			\end{tikzpicture}
	    	\end{adjustbox}}
	    	\subfigure[]{
	    		\begin{adjustbox}{width=0.5\textwidth}
	    			\begin{tikzpicture}[>=stealth]
	    				\node at (0,0) [circle,draw=black!100,inner sep=2.2mm] {\Large9};
	    				\node at (0,3) [circle,draw=black!100,inner sep=2.2mm] {\Large8};
	    				\node at (3,3) [circle,draw=black!100,inner sep=1.5mm] {\Large11};
	    				\node at (6,3) [circle,draw=black!100,inner sep=1.5mm] {\Large14};
	    				\node at (-3,3) [circle,draw=black!100,inner sep=2.2mm] {\Large5};
	    				\node at (-6,3) [circle,draw=black!100,inner sep=2.2mm] {\Large3};
	    				\node at (-6,6) [circle,draw=black!100,inner sep=2.2mm] {\Large2};
	    				\node at (-6,9) [circle,draw=black!100,inner sep=2.2mm] {\Large1};
	    				\node at (-9,9) [circle,draw=black!100,inner sep=2.2mm] {\Large0};
	    				\node at (-3,9) [circle,draw=black!100,inner sep=2.2mm] {\Large4};
	    				\node at (0,9) [circle,draw=black!100,inner sep=2.2mm] {\Large7};
	    				\node at (3,9) [circle,draw=black!100,inner sep=1.5mm] {\Large10};
	    				\node at (6,9) [circle,draw=black!100,inner sep=1.5mm] {\Large12};
	    				\node at (6,6) [circle,draw=black!100,inner sep=1.5mm] {\Large13};
	    				\node at (9,9) [circle,draw=black!100,inner sep=1.5mm] {\Large15};
	    				\node at (0,12) [circle,draw=black!100,inner sep=2.2mm] {\Large6};
	    				\draw [thick] (0.5,3) -- (2.5,3);
	    				\draw [thick] (3.5,3) -- (5.5,3);
	    				\draw [thick] (-0.5,3) -- (-2.5,3);
	    				\draw [thick] (-3.5,3) -- (-5.5,3);
	    				\draw [thick] (0.5,9) -- (2.5,9);
	    				\draw [thick] (3.5,9) -- (5.5,9);
	    				\draw [thick] (6.5,9) -- (8.5,9);
	    				\draw [thick] (-0.5,9) -- (-2.5,9);
	    				\draw [thick] (-3.5,9) -- (-5.5,9);
	    				\draw [thick] (-6.5,9) -- (-8.5,9);
	    				\draw [thick] (0,0.5) -- (0,2.5);
	    				\draw [thick] (0,9.5) -- (0,11.5);
	    				\draw [thick] (6,3.5) -- (6,5.5);
	    				\draw [thick] (6,6.5) -- (6,8.5);
	    				\draw [thick] (-6,3.5) -- (-6,5.5);
	    				\draw [thick] (-6,6.5) -- (-6,8.5);
	    				\draw (0.8,1.5) node {\large 1.045e-2};
	    				\draw (6.8,4.5) node {\large 8.800e-3};
	    				\draw (6.8,7.5) node {\large 6.825e-3};
	    				\draw (0.8,10.5) node {\large 1.073e-2};
	    				\draw (-5.2,4.5) node {\large 1.332e-2};
	    				\draw (-5.2,7.5) node {\large 1.208e-2};
	    				\draw (1.5,3.25) node {\large 9.076e-3};
	    				\draw (4.5,3.25) node {\large 7.613e-3};
	    				\draw (1.5,9.25) node {\large 1.523e-2};
	    				\draw (4.5,9.25) node {\large 1.326e-2};
	    				\draw (7.5,9.25) node {\large 5.464e-3};
	    				\draw (-1.5,3.25) node {\large 7.481e-3};
	    				\draw (-4.5,3.25) node {\large 1.187e-2};
	    				\draw (-1.5,9.25) node {\large 2.458e-2};
	    				\draw (-4.5,9.25) node {\large 8.158e-3};
	    				\draw (-7.5,9.25) node {\large 1.206e-2};
	    			\end{tikzpicture}
	    	\end{adjustbox}}
	    	\caption{Two coupling graphs of the IBMQ topology architecture, where each node represents a physical qubit and each edge represents that there is an interaction between the two qubits, and the values adhering to the edges are the error rates of the CNOT gates. (a) The IBMQ\_Quito topology with 5 qubits. (b) The IBMQ\_Guadalupe topology with 16 qubits.}
	    	\label{fig2}
	    \end{figure}
	    
	    Another physical constraint reflected in the coupling graph is the gate error. The weights on the edges of the topological graph indicate the error rates of gate operations between two qubits. Higher weights imply higher error rates. Additionally, due to the qubit quality parameter, the error rate between two adjacent qubits is also different, which leads to different error rates for CNOT gates acting on different qubits. Therefore, it is necessary to satisfy the connectivity constraint and to reduce the gate error when mapping quantum circuits.
	    
	    \subsection{Boolean matrix}
\begin{figure}[]
        	 \includegraphics[width=0.48\textwidth]{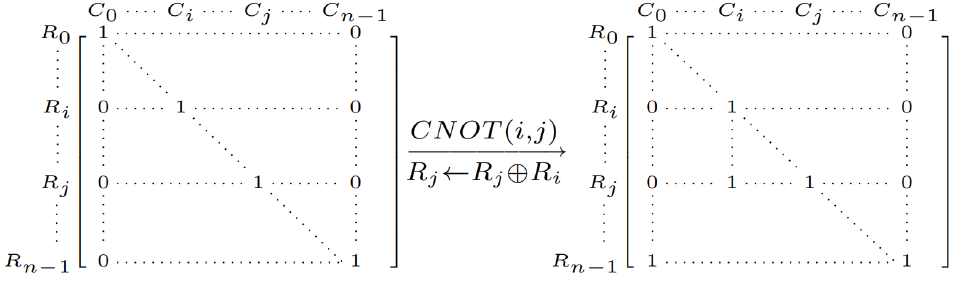}
        \caption{ To calculate the Boolean matrix of a CNOT circuit, the following steps can be followed. For any CNOT gate with $n$ variables in the circuit, where the variable domain is $\{x_1, x_2, ..., x_i, ..., x_j, ..., x_n\}$, with $x_i$ as the control bit and $x_j$ as the target bit. In an $n \times n$ identity matrix, the row $i$ corresponding to the control bit $x_i$ of each CNOT gate is XORed to the row $j$ corresponding to the target bit $x_j$. }
        \label{genelcnot}
        \end{figure}

	    The CNOT gate can be represented by a unitary matrix of size $4 \times 4$. The matrix of a CNOT circuit containing $n$ qubits requires a tensor product of the individual CNOT's unitary matrix, so a CNOT circuit with $n$ qubits requires a matrix representation of size $2^n \times 2^n$. However, in the Boolean matrix representation of a CNOT circuit, the matrix size corresponds to the number of qubits. A CNOT circuit with $n$ number of qubits corresponds to an $n \times n$ Boolean matrix.  The Boolean matrix of the CNOT circuit is calculated as shown in Fig.~\ref{genelcnot}.

	    \begin{figure}
	    		\centering
	    		\subfigure[]{
	    			\begin{adjustbox}{width=0.35\textwidth}
	    				\begin{quantikz}[row sep={0.6cm,between origins}]
	    					\lstick{$q_0$} &  \qw    &  \qw  &   \ctrl{4} &  \qw       &  \ctrl{2}  &  \qw   \\		
	    					\lstick{$q_1$} &  \qw    &  \qw  &  \qw       &   \ctrl{2} &  \qw  &  \qw    \\
	    					\lstick{$q_2$} &  \targ{}\gategroup[wires=1,steps=1,style={opacity=0,rounded corners, inner sep=1pt}, label style={label position=below, yshift=-0.15cm}]{\footnotesize }   & \ctrl{2}\gategroup[wires=1,steps=1,style={opacity=0,rounded corners, inner sep=1pt}, label style={label position=below, yshift=-0.2cm}]{\footnotesize } &  \qw      &  \qw       &  \targ{}\gategroup[wires=1,steps=1,style={opacity=0,rounded corners, inner sep=1pt}, label style={label position=below, yshift=-0.2cm}]{\footnotesize $G_5$}  &  \qw   \\
	    					\lstick{$q_3$}  & \ctrl{-1}\gategroup[wires=1,steps=1,style={opacity=0,rounded corners, inner sep=1pt}, label style={label position=below, yshift=-0.15cm}]{\footnotesize $G_1$}  &   \qw      &  \qw      &  \targ{}\gategroup[wires=1,steps=1,style={opacity=0,rounded corners, inner sep=1pt}, label style={label position=below, yshift=-0.2cm}]{\footnotesize $G_4$}  &  \qw    &  \qw    \\
	    					\lstick{$q_4$}  &   \qw   &  \targ{}\gategroup[wires=1,steps=1,style={opacity=0,rounded corners, inner sep=1pt}, label style={label position=below, yshift=-0.2cm}]{\footnotesize $G_2$}    &  \targ{}\gategroup[wires=1,steps=1,style={opacity=0,rounded corners, inner sep=1pt}, label style={label position=below, yshift=-0.2cm}]{\footnotesize $G_3$}     &  \qw      &  \qw &  \qw     
	    				\end{quantikz}
	    		\end{adjustbox}}
	    		\subfigure[]{
	    				\begin{adjustbox}{width=0.35\textwidth}
	    					\centering
	    					\includegraphics{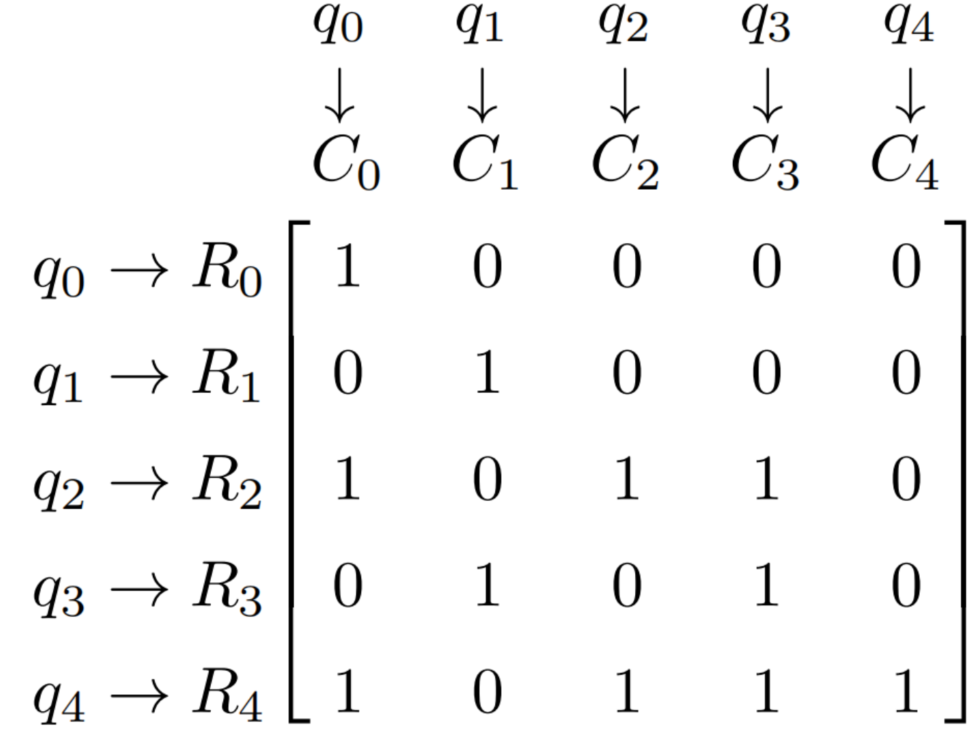}
	    				\end{adjustbox}
	    		}
	    
	    	\caption{A CNOT circuit and its Boolean matrix. (a) A CNOT circuit consisting of 7 CNOT gates. (b) Correspondence between the rows and columns of the Boolean matrix and the qubits.}
	    	\label{fig3}
	    \end{figure}
	    
	    The CNOT circuit shown in Fig.~\ref{fig3}(a) consists of seven CNOT gates with five qubits, namely $\{q_0,q_1,q_2,q_3,q_4\}$. It can be represented as a $5 \times 5$ Boolean matrix, where $\{q_0,q_1,q_2,q_3,q_4\}$ correspond to $\{R_1,R_2,R_3,R_4,R_5\}$ and $\{C_1,C_2,C_3,C_4,C_5\}$ in the Boolean matrix, respectively. During the conversion process of Boolean matrix, several XOR operations are performed. Firstly, the control bit $q_3$ of gate $G_1$ is XORed to the target bit $q_2$ and the result is stored in the target bit $q_2$, which is expressed as $R_2=R_2 \oplus R_3$ in the Boolean matrix. And then the control bit $q_2$ of gate $G_2$ is XORed to the target bit $q_3$, which is expressed as $R_3=R_3 \oplus R_2$ in the Boolean matrix. The same operation is carried out for $G_2$ to $G_5$. The final Boolean matrix is shown in Fig.~\ref{fig3}(b).

	\section{Key-qubit priority initial mapping strategy}
	\label{sec:keyqubit}
	In order to extend the synthesis method with Hamiltonian path architectures, and solve the NN synthesis problem of CNOT circuits on the architecture without Hamiltonian paths, the nearest-neighbor synthesis of CNOT circuits on general quantum architectures is proposed to achieve the NN synthesis of CNOT circuits while ensuring the NN constraint. The NN synthesis of CNOT circuits on general quantum architectures is ultimately achieved through a noise-aware CNOT circuit NN synthesis strategy (Algorithm~\ref{algo4}). This algorithm comprises two main components: initial mapping and matrix elimination. The initial mapping is conducted using a key-qubit priority initial mapping strategy (Algorithm~\ref{algo1}) along with its optimized variant (Algorithm~\ref{KOPIMO}). Matrix elimination is implemented using a Steiner Tree approach and target-aided rows matching (Algorithm~\ref{algo3}).
 
    This section proposes a key-qubit priority mapping model that prioritizes and maps key qubits on a coupling graph. Additionally, tabu search is employed to optimize the initial mapping, aiming to reduce the number of CNOT gates and enhance the circuit fidelity.

	\subsection{Key-qubit priority model}
\begin{figure}
	\centering
	\begin{adjustbox}{width=0.45\textwidth}
		\includegraphics{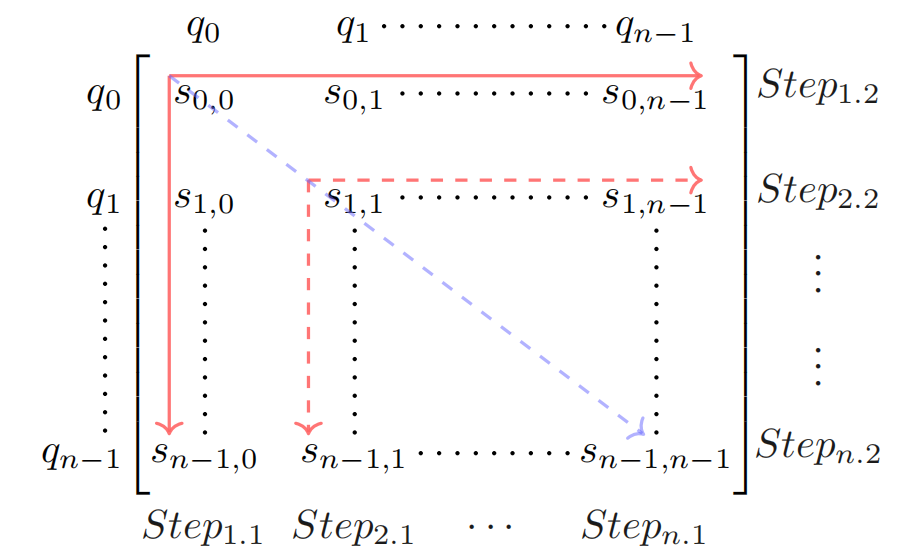}
	\end{adjustbox}
	\caption{Illustration of the Boolean matrix elimination process. The Boolean matrix is processed by layer and follows the principle of columns before rows.}
	\label{figa4}
    \end{figure}

      An \(n \times n\) Boolean matrix can be divided into \(n\) layers along its diagonals, where the first layer includes the first column and the first row, the \(i\)-th layer (\(1 \leq i \leq n\)) includes the \(i\)-th column and \(i\)-th row, up to the last layer which contains only one element. In the process of Boolean matrix elimination, a layer-first principle is followed, as shown in Fig.~\ref{figa4}. The layer-first principle starts from the first layer and proceeds to the (\(n-1\))-th layer, and the elimination is carried out in the order of columns first and then rows. Ultimately, the Boolean matrix is transformed into an identity matrix, and the synthesized quantum circuit comprises a reverse cascade of all the CNOT gates applied throughout the transformation process.
     
    Given that each layer of the Boolean matrix represents a qubit, processing the matrix layer by layer corresponds to the operation sequence $\{q_0, q_1, \ldots, q_n\}$. The corresponding operation sequence in the initial mapping of the coupling graph is $\{q_0 \rightarrow Q_{i0}, q_1 \rightarrow Q_{i1}, \ldots, q_n \rightarrow Q_{in}\}$. At every $i$-th step, it is necessary to remove vertex $Q_i$, ensuring that the diagonal elements of the matrix, post-layer operation, are not affected by subsequent synthesis processes. However, if the operation sequence in the coupling graph post-initial mapping does not form a Hamiltonian path, the step of removing vertices may result in the coupling graph being divided into two disconnected subgraphs. This division causes a loss of interaction paths between certain qubits, rendering the circuit unsynthesized. Therefore, it is crucial to find a reasonable initial mapping method that prevents the removal of vertices from dividing the coupling graph into disconnected subgraphs.

	\begin{lemma}
		Let $G = (V,E)$ be an undirected graph. If the removal of a vertex $v \in V$, along with all the edges associated with it, causes $G$ to split into two or more disconnected subgraphs, then the vertex $v$ is known as the cut point of graph $G$~\cite{nadler1993continuum}.
	\end{lemma}
	
	In the coupling graph shown in Fig.~\ref{fig4}, $Q_1$ and $Q_3$ are cut points. Their deletion from the coupling graph results in the splitting of the graph into multiple subgraphs, rendering the matrix elimination process infeasible. In order to perform matrix elimination operations accurately, it is necessary to map logical qubits with lower indexes to vertices that are not cut points in the physical quantum coupling graph in priority during the initial mapping. We refer to such vertices as key qubits.
	
	\begin{definition}
		For a coupling graph $G=(Q,E)$ that contains several physical qubits, where $Q= \{Q_0,Q_1,\dots,Q_i,\dots,Q_{n-1}\}$ $(0 \leq i~\textless\ n)$, a vertex $Q_i$ that is a non-cut point is regarded as the key physical qubit of the priority mapping, or the key qubit for brevity.
	\end{definition}
	
	To ensure the feasibility of matrix elimination, priority is given to the physical qubits that are non-cut points, known as key qubits, during the mapping process. For instance, in the IBMQ\_Quito architecture, qubits $Q_0$, $Q_2$, and $Q_4$ are non-cutting points, which are the key qubits, as shown in Fig.~\ref{fig4}. During the initial mapping, priority is given to map $q_0$, $q_1$, $q_2$ to these three physical qubits. If $q_0$ is the first qubit to be mapped to $Q_0$, it becomes disconnected from the other qubits in the coupling graph after the first layer of matrix elimination is completed. By prioritizing the constraints of mapping key qubits, the key-qubit priority model can be satisfied while reducing the complexity of the initial mapping at the same time.
	
	\begin{figure}
		\centering
		\begin{adjustbox}{width=0.25\textwidth}
			\begin{tikzpicture}[>=stealth]
				\node (q1) at (-6,3) [circle,dashed,draw=black!100,inner sep=2.2mm] {\LARGE $Q_4$};
				\node (q2) at (-6,6) [circle,draw=black!100,inner sep=2.2mm] {\LARGE $Q_3$};
				\node (q3) at (-6,9) [circle,draw=black!100,inner sep=2.2mm] {\LARGE $Q_1$};
				\node (q0) at (-9,9) [circle,dashed,draw=black!100,inner sep=2.2mm] {\LARGE $Q_0$};
				\node (q4) at (-3,9) [circle,dashed,draw=black!100,inner sep=2.2mm] {\LARGE $Q_2$};
				\draw [thick] (q0) to (q3);
				\draw [thick] (q3) to (q4);
				\draw [thick] (q2) to (q3);
				\draw [thick] (q1) to (q2);
				\draw [line width = 2pt] (-7.7,9.3) -- (-7.2,8.7);
				\draw [line width = 2pt] (-7.2,9.3) -- (-7.7,8.7);
			\end{tikzpicture}
		\end{adjustbox}
		\caption{An example of the key-qubit priority model for the IBMQ\_Quito. The dashed qubits $Q_0$, $Q_2$, and $Q_4$ are non-cutting points, i.e., key qubits. These qubits need to be prioritized. Solid qubits $Q_1$ and  $Q_3$ are cut points.}
		\label{fig4}
	\end{figure}
	
		\subsection{Key-qubit priority mapping based on tabu search}
	In order to avoid the occurrence of disconnected subgraphs in a coupling graph, this paper proposes a key-qubit priority mapping method.  This method prioritizes the vertices that are not cut points, i.e., key points, on the coupling graph. Then the processed vertices and the edges connected to them are removed from the coupling graph until a complete Hamiltonian path exists in the updated coupling graph. After the deletion of a key qubit, a new subgraph is created. New key qubits may appear in this subgraph, which were not cut points in the original coupling graph. By considering the constraints of key qubits, a recursive algorithm for selecting and mapping the key qubits in the order of increasing logical qubits index can be provided in the analysis of the key-qubit priority model. This ensures that the logical qubits at the key qubits are prioritized in the subsequent elimination process. The key-qubit priority initial mapping algorithm KQPIM is shown in Algorithm \ref{algo1}.
	
	\begin{algorithm}[]
		\KwIn{The coupling graph $G=(V,E)$, the initial map $\pi$, the number of logical qubits $n$, the list of initial key qubits $ikey\_list$}
		\KwOut{The initial map $\pi_0$}
		\Begin{
			$i \leftarrow $ Nodes($G_0$) - Nodes($G$);\\
			\If{$G  \ \exists \  Hamilton \ path$}{
				\ForEach{$Q_i$ \rm in $Hamilton \ path$}{
					add \{$i:Q_i$\} to $\pi$;\\
					$i++$;
				}	
				\Return $\pi$;
			}
			\If{$\pi$ = null}{
				$key\_list \leftarrow ikey\_list[0]$ ;\\
				$G \leftarrow G/ikey\_list[0]$; \tcc*[f]{Delete point from $G$}\\
			}
			\ForEach{$Q$ \rm in $V$}{
				$key\_list \leftarrow Q$ is non-cut points;\\
			}	
			\If{$G \ \exists\mkern-10mu/ \ Hamilton \ path$ }{
				$key\_qubit \leftarrow key\_list[random]$;\\
				add \{$i:key\_qubit$\} to $\pi$;\\
				$G \leftarrow G/key\_qubit$; \\
				$i++$;\\
			}
			\If{\rm len($\pi$) != $n$}{
				KQPIM($G,\pi,n,ikey\_list$);\\
			}
		}
		\caption{Key-qubit priority initial mapping (KQPIM)}
		\label{algo1}
	\end{algorithm}

	This algorithm takes the coupling graph $G$ and the number of logical qubits $n$ as input, and outputs a randomized key-qubit priority initial mapping. This algorithm takes the non-cut points in the coupling graph as key qubits and iterates through recursion, reducing the complexity of the initial mapping. However, it is worth noting that Algorithm~\ref{algo1} only outputs a single mapping scheme. With $n$ qubits in a quantum computing architecture, there can be up to $n!$ possible initial mapping schemes, making it impractical to traverse each one by the brute force search method. Although the key-qubit priority initial mapping algorithm can reduce the search space of initial mappings, it may still not be able to find the optimal initial mapping quickly considering the increasing size of physical qubits on quantum computing devices. In this case, a better balance between solution accuracy and speed can be achieved with the help of metaheuristic algorithms.
	
	This work utilizes a tabu search algorithm to solve the initial mapping problem for CNOT circuit synthesis. Tabu search guides the search toward a more optimal region of the search space by maintaining a tabu table, while avoiding exploring suboptimal solutions previously~\cite{glover1993user}. The algorithm iterates and updates the tabu table as it explores the solution space, gradually improving the quality of the found solution until a satisfactory solution is found or the maximum number of iterations is reached. Since the elimination of the Boolean matrix under topological constraints requires considering not only the CNOT cost under the NN constraint but also the overall fidelity of the circuit. The number of CNOT gates is highly correlated with the connectivity of the qubits in the coupling graph. Low qubits connectivity can necessitate more CNOT gates to complete the NN. On the other hand, the error rate of CNOT gates is a crucial factor affecting the fidelity of the circuit. Choosing qubits with a lower error rate for mapping is beneficial for building circuits with high fidelity. The objective function $F_\pi$ of the optimization method is defined in Eq.~\eqref{eq4},
	\begin{equation}
		F_\pi=\prod\limits_{0 \leq i<j<n} \frac{\sum\limits_{\substack{v \neq i, j}} \frac{C_v(i, j)}{C_v}}{C_{ij}}-\sum\limits^{n-1}_{m=0}{(m+1){\frac{\sum\limits_{\substack{w\in E(\pi_m)}} w}{\left| E(\pi_m) \right|}}}
		\label{eq4}
	\end{equation}
	where $n$ denotes the number of qubits, $C_v (i,j)$ denotes the number of shortest paths through vertex $v$ that contain both vertices $i$ and $j$, $C_v$ denotes the number of shortest paths through vertex $v$, $C_{ij}$  denotes the number of shortest paths through $i$ and $j$. $m$ denotes the index number in the mapping method $\pi$, $\pi_m$ denotes the $m$-th physical qubit in the mapping method $\pi$, $E(\pi_m)$ denotes the edge connected to the qubit $\pi_m$ in the coupling graph, and $w$ denotes the weight on this edge, i.e., the error rate. 
	
	The first half of Eq.~\eqref{eq4} represents the degree of connectivity of the mapped qubits in terms of connectivity factors. This part is used to measure the connectivity of the sub-coupling graph when the physical qubits are not fully mapped. The inter-vertex connectivity factor is one of the indicators of the connectivity between vertices in a graph. The connectivity factor can be calculated by using the Betweenness Centrality algorithm in graph theory~\cite{brandes2001faster}. It ranges from 0 to 1, where 0 indicates that two vertices are not connected. A higher connectivity factor indicates stronger connectivity and fewer CNOT gates required to act on the two qubits. 
	And the second half of Eq.~\eqref{eq4} represents the fidelity profile of the mapped qubits. According to the key-qubit priority model, the more advanced qubits in $\pi$ are removed first, and the error rate associated with these qubits also have less impact on the circuit. To distinguish the importance of the error rates between different qubits, $m+1$ is assigned as the weight of these error rates. The later the qubits in $\pi$ are removed, i.e., the larger $m$ is, the greater the number of CNOT gates acting on these qubits, hence, the greater the impact on circuit fidelity. Therefore, the weights of these qubits are set through $m+1$.
	
	The objective function Eq.~\eqref{eq4} directly addresses the dual objectives of optimizing qubit connectivity and minimizing error rates to enhance the overall fidelity of the quantum circuit. High connectivity among qubits typically suggests fewer CNOT gates are required, potentially reducing the circuit's overall execution time and complexity. However, high connectivity does not automatically guarantee high fidelity; the error rates associated with the qubits and their connections play a critical role. High error rates can significantly degrade the circuit's fidelity, negating the benefits of high connectivity.	Therefore, the objective function Eq.~\eqref{eq4} is designed to balance these considerations by weighing both the connectivity of qubits and their fidelity profiles. This balanced approach allows for the identification of qubit mappings that not only ensure efficient quantum gate implementations but also maintain high circuit fidelity. By simultaneously optimizing for connectivity and error rates, Eq.~\eqref{eq4} serves as a comprehensive cost function that guides the tabu search algorithm towards finding mappings that achieve high-fidelity quantum circuit designs under practical constraints.
	
	After the cost function has been determined, the initial mapping of CNOT circuit synthesis is optimized by tabu search algorithm. Firstly, a tabu table of length $N$ is initialized, and the mapping method $\pi_0$ of Algorithm 1 (KQPIM) is added to the tabu table. Secondly, a set of candidate solutions is generated, namely multiple mapping methods $\pi_{\Delta} = \{\pi_1,\pi_2,\dots,\pi_n\}$, by randomly disturbing the initial key qubits in $\pi_0$. These initial key qubits are the non-cut points of the uncut coupling graph, and these qubits form the ikey\_list. Disturbing only the ikey\_list ensures that the key qubit priority model is satisfied. Next, the mapping schemes in $\pi_{\Delta}$ that already exist in the tabu table are removed, and the mapping schemes whose cost $F_{\pi}$ is lower than the average cost in the tabu table are added to the tabu table. Finally, the tabu table is updated to remove the mappings with higher cost $F_{\pi}$ in the tabu table. This cycle continues until the end of the iteration. The pseudo-code for the tabu search algorithm to optimize the initial mapping of CNOT circuit synthesis is provided in Algorithm~\ref{KOPIMO}.

	\begin{algorithm}[]
		\SetKwComment{Comment}{start}{end}
		\KwIn{The coupling graph $G=(V,E)$, the number of logical qubits $n$}
		\KwOut{The best initial map $\pi_{best}$}
		\Begin{
			$T\_list \leftarrow [\ ]*N$, $\pi\leftarrow [\ ]$;\\
			$ikey\_list \leftarrow$ non-cut points in $G$;\\
			$\pi_0 \leftarrow KQPIM(G,\pi,n,ikey\_list)$;\\
			add $\pi_0$ to $T\_list$;\\
			\For{$i < $ iterations}{
				$\pi_\Delta \leftarrow [\ ]$ ;\\
				\For{$k = 0$ \rm to $N$}{
					\tcp{Change only key qubits}
					$ikey\_list \leftarrow$ random disturbance key qubits in $ikey\_list$;\\
					$\pi_k \leftarrow$ $KQPIM(G,\pi,n,ikey\_list$);\\
					add $\pi_k$ to $\pi_\Delta$;	\\
				}
				\ForEach{$\pi$ \rm in $\pi_\Delta$ }{
					
					\If{$\pi \notin T\_list$ \rm and $F_\pi \leq \rm F_{avg(T\_list)}$ }{
						add $\pi$ to $T\_list$;\\
					}
					\If{size$(T\_list) >$ N}{$T\_list \leftarrow$ update($T\_list$)}
				}
			}
			$\pi_{best} \leftarrow $ best($T\_list$);\\
			\Return $\pi_{best}$;
		}
		\caption{Key-qubit priority initial mapping optimization (KQPIMO)}
		\label{KOPIMO}
	\end{algorithm}
	
	The key-qubit priority initial mapping optimization algorithm based on tabu search described above is an iterative optimization method. It helps to improve the optimization rate and fidelity of quantum circuits by setting a reasonable length of the tabu table and the number of iterations. This enables the mapping method to trade off a certain amount of time cost for the enhancement of the overall quality of the circuit.
	
	\section{Key qubit-aware CNOT circuit NN synthesis}
	\label{sec:lcnns}
	This section presents an NN synthesis method based on the key-qubit priority initial mapping strategy. The method improves the synthesis efficiency by prioritizing the mapping and elimination processing of key qubits. It enables the implementation of CNOT circuit NN synthesis on an architecture with and without Hamiltonian paths. And at the same time, the fidelity of the NN synthesized circuits is improved as much as possible. Specifically, the target-aided rows matching algorithm is introduced firstly, which helps to find the best auxiliary row to achieve row elimination of Boolean matrix. Next, this section describes the noise-aware CNOT circuit NN synthesis algorithm based on layer convergence, which uses a layer-by-layer elimination method. Finally, an algorithm example is provided to facilitate a better understanding of the application and effect of the algorithm.
	\subsection{Target-aided rows matching algorithm}
	In the key-qubit priority mapping model, the processed vertices need to be removed from the coupling graph. Consequently, the elimination must process not only the current column but also the current row. In order to change the values of the elements in the current row other than those on the main diagonal to 0, it is necessary to use the auxiliary of other rows in the matrix to perform an XOR operation with the current row. The set of target-aided rows found based on the current row is presented in Definition~\ref{df3}.,
	
	\begin{definition}
        Consider an invertible Boolean matrix. Let \(R_i\) be the current row and \(e_i\) be its corresponding unit row vector, defined such that \(e_{ij} = 1\) if \(j = i\) and \(e_{ij} = 0\) otherwise. Define the set \(Set_k\) comprising rows \(R_k\) for which the \(\bigoplus R_k = R_i \oplus e_i\) holds, where \(i < k \leq n\) and \(n\) is the number of rows in the matrix. This set is referred to as the target-aided rows set.
		\label{df3}
	\end{definition}
	
\begin{figure}[]
		\centering  
		\begin{adjustbox}{width=0.45\textwidth}
			\includegraphics{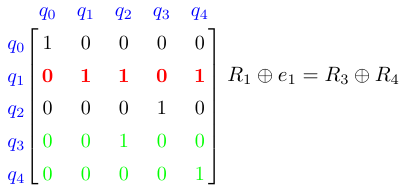}
		\end{adjustbox}
		\caption{Target-aided rows matching example. By calculating $R_1 \oplus e_1=[0,0,1,0,1]$, the set of target-aided rows must be matched to obtain  $\bigoplus R_j =[0,0,1,0,1]$. And the set \{$R_3,R_4$\} satisfies $R_3\oplus R_4=[0,0,1,0,1]$, therefore, the target-aided rows set of $R_1$ is \{$R_3,R_4$\}.}
		\label{fig7}
	\end{figure}

	\begin{example}
		Fig.~\ref{fig7} illustrates an example of matching the target-aided rows. At this point, the rows of the first layer and the columns of the second layer of this matrix have been processed, and the next step is to complete the row elimination of the second layer. According to Definition~\ref{df3}, it is necessary to match to the set of target-aided rows that satisfy the condition $\bigoplus R_j=R_1 \oplus e_1$, where $e_1=[0,1,0,0,0]$. After calculating, the target-aided rows set of $R_1$ is \{$R_3,R_4$\}.
	\end{example}
	
	In the case where the matrix is invertible, the system of linear equations $Ax = y$ must have a unique solution. Therefore, if the target row exists, the set of target-aided rows must be found, i.e., the target-aided rows matching is completed. Specifically, the set of target rows can be determined by solving the system of linear equations. If the system of linear equations has a solution, it means that the target rows set exists, and we only need to find the rows corresponding to the non-zero elements in the solution vector. Based on the above analysis, a target-aided rows matching algorithm is given, as shown in Algorithm~\ref{algo3}.
	
	\begin{algorithm}[]
		\SetKwComment{Comment}{start}{end}
		\KwIn{The invertible boolean matrix $M$, the target row index $i$}
		\KwOut{The target auxiliary row index list $R_j$}
		\Begin{
			\tcp{Get the rank of the matrix}
			$n \leftarrow r(M)$;  \\
			\tcp{Initialize list to generate permutations} 
			$L \leftarrow [\ ], R\_list \leftarrow [\ ]$; 	\\
			\For{$m = i$ \rm to $n$}{
				\tcp{Generate the index sequence to be matched according to the target row index}
				add $m$ to $L$;	\\
			}	
			\For{$length = i$ to $n$}{
				\tcp{save current permutation}
				$R\_list \leftarrow$ Prem.append($L,length$); \\
			}
			\ForEach{$R_j$ \rm in $R\_list$}{
				\If{$\bigoplus R_j = R_i \oplus e_i$}{\Return $R_j$;}
			}
		}
		\caption{Target-aided rows matching (TARM)}
		\label{algo3}
	\end{algorithm}
	
	To construct the set of target-aided rows, the algorithm first generates the sequence of indexes to be matched in accordance with the target row index. Next, the algorithm searches for feasible solutions through the full permutations of all rows, except for the target row. The full permutation of $l$ numbers from the set $L$ can be expressed as Perm ($L,l$). So the complexity of lines 7-8 in Algorithm~\ref{algo3} is shown in Eq.~\eqref{eq7}.
	\begin{equation}
		O\left(\sum_{l=0}^n \frac{n !}{(n-l) !}\right)=O\left(n ! \sum_{l=0}^n \frac{1}{l !}\right)
		\label{eq7}
	\end{equation}
	where $n$ denotes the number of qubits. The value of $\sum_{l=0}^n \frac{1}{l !}$ in Eq.~\eqref{eq7} must be less than 3, and it is a constant. In addition, the complexity of the summation operation in line 11 of Algorithm~\ref{algo3} is $O(n)$, and this operation is nested within $O(n!)$. Therefore, the complexity of Algorithm~\ref{algo3} is $O(n\cdot n!)$.
	
\subsection{Noise-aware CNOT circuits NN synthesis strategy}
	
    The synthesis of CNOT circuits can be accomplished by using the Steiner tree model and Gaussian elimination to convert the Boolean matrix $M$ of the CNOT circuit to an identity matrix $I$. Within the quantum coupling graph, each column of the Boolean matrix with qubits represented by the value 1 can be used to construct a Steiner tree. This Steiner tree includes all qubits (terminal nodes) with a value of 1 as well as auxiliary qubits (Steiner nodes). The construction of the Steiner tree has been proven to be an NP-complete problem~\cite{karp2010reducibility}. Subsequent row operations in Gaussian elimination are constrained by the edges of the Steiner tree, thus ensuring that the corresponding CNOT gates adhere to the connectivity constraints of the architecture.

    \begin{definition}
In Gaussian elimination, a CNOT gate modifies matrix elements. To switch element \(S_i\) from 0 to 1 using \(S_j=1\), apply \(S_i \leftarrow S_i \oplus S_j\) by a CNOT gate CNOT($S_j,S_i$), denoted as \(\text{set1}[S_i, S_j]\), achieving \(0 \oplus 1 = 1\). Similarly, to set \(S_i\) from 1 to 0, use \(S_i \leftarrow S_i \oplus S_j\), noted as \(\text{set0}[S_i, S_j]\), resulting in \(1 \oplus 1 = 0\).
	\end{definition}

    Based on the preceding elimination rules, the Boolean matrix $M$ must be processed sequentially by column. A Steiner tree is constructed for each  $i$-th column of the matrix. The $set1$ or $set0$ operation is then employed to convert all non-zero elements in each column to 0, except for the $i$-th element, which is transformed to 1. 
    
 	\begin{example}
		Taking the circuit in Fig.~\ref{fig3}(a) as an example, the circuit is mapped onto the coupling graph in Fig.~\ref{fig2}(a) in the default initial mapping \{$q_0 \rightarrow Q_0, q_1 \rightarrow Q_1, q_2 \rightarrow Q_2,q_3 \rightarrow Q_3,q_4 \rightarrow Q_4$\}. The first column of the Boolean matrix $M$, which represents this circuit, is processed first. A Steiner tree is constructed for the elements of the first column. In this Steiner tree, $q_0$ acts as the root node, and $q_2$ and $q_4$ are the leaf nodes. The Steiner tree is then post order traversed, and the sequence of qubits elimination is $q_4$, $q_2$, $q_0$. While eliminating $q_4$, it is not possible to convert $q_4$ to 0 by using $set0(q_4, q_3)$ because the state of $q_3$, which is adjacent to $q_4$, is 0. Therefore, $set1(q_3,q_4)$ is used to convert $q_3$ to 1, followed by the execution of $set0(q_4, q_3)$. The corresponding CNOT operations are CNOT(4,3) and CNOT(3,4). Similarly, the same operation is performed on the other qubits until only $q_0$ is 1 in the first column of this matrix. The process of synthesis of the first column is shown in Fig.~\ref{NNsynthesis}. The numbers indicate the order of the set operations, the solid arrows indicate $set1$, the dashed arrows indicate $set0$.
	\end{example}
 
		\begin{figure}
		\centering
		\begin{adjustbox}{width=0.25\textwidth}
			\begin{tikzpicture}[>=stealth]
				\node (q4) at (-6,3) [circle,draw=black!100,inner sep=2.2mm] {\LARGE $q_4$};
				\node (q3) at (-6,6) [circle,draw=black!100,inner sep=2.2mm] {\LARGE $q_3$};
				\node (q1) at (-6,9) [circle,draw=black!100,inner sep=2.2mm] {\LARGE $q_1$};
				\node (q0) at (-9,9) [circle,draw=black!100,inner sep=2.2mm] {\LARGE $q_0$};
				\node (q2) at (-3,9) [circle,draw=black!100,inner sep=2.2mm] {\LARGE $q_2$};
				\draw [thick] (q0) to (q1);
				\draw [thick] (q3) to (q4);
				\draw [thick] (q2) to (q1);
				\draw [thick] (q1) to (q3);
				\draw[-latex,dashed] (q0) to[out = 45,in=135]  (q1);
				\draw[-latex,dashed] (q1) to[out = 45,in=135]  (q2);
				\draw[-latex,dashed] (q1) to[out = 225,in=135]  (q3);
				\draw[-latex,dashed] (q3) to[out = 225,in=135]  (q4);
				\draw[-latex] (q3) to[out = 45,in=315]  (q1);
				\draw[-latex] (q4) to[out = 45,in=315]  (q3);
				\draw (-4.8,4.5) node {\large{\textcircled{\scriptsize{1}}}\large};
				\draw (-4.8,7.5) node {\large{\textcircled{\scriptsize{2}}}\large};
				\draw (-7.2,4.5) node {\large{\textcircled{\scriptsize{3}}}\large};
				\draw (-7.2,7.5) node {\large{\textcircled{\scriptsize{4}}}\large};
				\draw (-4.5,10.1) node {\large{\textcircled{\scriptsize{5}}}\large};
				\draw (-7.5,10.1) node {\large{\textcircled{\scriptsize{6}}}\large};
			\end{tikzpicture}
		\end{adjustbox}
		\caption{The NN synthesis process in the IBMQ\_Quito architecture for the circuit of Fig.~\ref{fig3}(a). The numbering indicates the order of the sets, the dotted arrows indicate $set0$, the solid arrows indicate $set1$.}
		\label{NNsynthesis}
	\end{figure}
 
    Given that quantum algorithms involve many two-qubit operations, the error of each two-qubit operation, i.e., the two-qubit gate, is more significant compared to single-qubit operations. In order to maximize the probability of successful execution of the synthesized CNOT circuits on real quantum computing devices, the error of each inserted CNOT gate must be considered during the synthesis of CNOT circuits. A noise-aware CNOT circuit NN synthesis method is proposed to mitigate these unavoidable noise problems.
	
	In the process of matrix elimination, multiple interaction paths may occur when Gaussian elimination is performed between two qubits that are not close neighbors. Since the error rates between neighboring qubits on different paths are different, choosing different interaction paths to build Steiner trees can cause different errors in the results. To measure the error cost of each interaction path, we propose an interaction path fidelity measure $F_{path}$, as shown in Eq.~\eqref{eq666}.
	\begin{equation}
		F_{path}=\prod_{Q_i, Q_j \in p a t h}\left(1-E\left[Q_i\right]\left[Q_j\right]\right)
		\label{eq666}
	\end{equation}
	where $Q_i$ and $Q_j$ are the two qubits that are near neighbors on the interaction path, and $E\left[Q_i\right]\left[Q_j\right]$
	denotes the two-qubit gate operation error rate, i.e., the weight of the edge on the coupling graph. A larger $F_{path}$ indicates a higher fidelity on this interaction path, which means that the integrated CNOT quantum circuit can obtain a better reliability.

\begin{figure}[]
	\begin{adjustbox}{width=0.45\textwidth}
		\begin{tikzpicture}
			\begin{axis}[
				xlabel={},
				ylabel={\Large Fidelity},
				ymin=0.49, ymax=1.099,  
				width=15cm,
				height=.55\textwidth,
				legend pos=north west,
				legend style={at={(0.6,0.995)},anchor=north east,cells={anchor=west},draw = none,fill = none},
				xtick={1,2,3,...,15}, 
				xticklabels={
					4mod5-v1\_22, 4mod5-v1\_24, mod5mils\_65, alu-v0\_27, alu-v3\_35, alu-v4\_37, 
					4gt13\_92, 4mod5-v1\_23, decode24-v2\_43, 4gt5\_75, 4gt13\_91, alu-v4\_36, 
					4gt13\_90, hwb4\_49, mod10\_171
				}, 
				xticklabel style={rotate=30, anchor=north east}, 
				]
				\addplot[color=blue,mark=*, mark options={fill=blue}] coordinates {
					(1, 0.8661)
					(2, 0.7781)
					(3, 0.9052)
					(4, 0.9509)
					(5, 0.9333)
					(6, 0.8751)
					(7, 0.7713)
					(8, 0.8121)
					(9, 0.9815)
					(10, 0.7439)
					(11, 0.805)
					(12, 0.9057)
					(13, 0.7811)
					(14, 0.8398)
					(15, 0.9815)
				};
				\addplot[color=red,mark=square*, mark options={fill=red}] coordinates {
					(1, 0.8334)
					(2, 0.7087)
					(3, 0.8277)
					(4, 0.8664)
					(5, 0.8519)
					(6, 0.771)
					(7, 0.7412)
					(8, 0.7694)
					(9, 0.8729)
					(10, 0.7472)
					(11, 0.7597)
					(12, 0.8086)
					(13, 0.7742)
					(14, 0.8408)
					(15, 0.8696)
				};
				\addplot[color=black,mark=triangle*, mark options={fill=black},] coordinates {
					(1, 0.7431)
					(2, 0.61959)
					(3, 0.78039)
					(4, 0.9086)
					(5, 0.8734)
					(6, 0.8121)
					(7, 0.6548)
					(8, 0.69969)
					(9, 0.906)
					(10, 0.62189)
					(11, 0.6748)
					(12, 0.82999)
					(13, 0.666)
					(14, 0.7952)
					(15, 0.910099)
				};
				\addplot+[mark=none,color=blue,dashed] coordinates {
					(0.5,0.8) (15.5,0.93)
				};
				\addplot+[mark=none,color=red,dashed] coordinates {
					(0.5,0.73) (15.5,0.86)
				};
				\addplot+[mark=none,color=black,dashed] coordinates {
					(0.5,0.66) (15.5,0.79)
				};
				\legend{\large Fidelity(Simulation by Eq.~\eqref{eq666}),\large Fidelity(Actual),\large Fidelity(Simulation by fake\_provider)}
			\end{axis}
		\end{tikzpicture}
	\end{adjustbox}
	\caption{Actual fidelity versus simulated fidelity on IBMQ\_Quito. The red curve represents the fidelity achieved through execution on an actual quantum computing device IBMQ\_Quito, the black curve represents the fidelity of IBMQ\_Quito simulation using qiskit's fake\_provider simulator, and the blue curve corresponds to the fidelity simulated using Eq.~\eqref{eq666}. }
	\label{fig13vs}
\end{figure}

   	In order to verify the accuracy of the fidelity calculation method in Eq.~\eqref{eq666}, the fidelity calculated from the simulations is compared with the fidelity of the actual quantum hardware after implementation, and the results are shown in Fig.~\ref{fig13vs}. By comparing the fidelity of 15 circuits from RevLib \cite{wille2008revlib}, whose circuit sizes range from 11 to 108, it is found that the simulated fidelity shows essentially the same fluctuations and trends as the actual fidelity, but the simulated fidelity is slightly higher than the actual fidelity. This phenomenon can be attributed to the existence of some specific noise and errors in the actual quantum hardware, such as crosstalk, measurement errors, and some other quality parameters of the qubits that may change over time. These factors worsen the error rate and lower the actual fidelity of the synthesized CNOT circuits. Similarly, the fidelity of simulated circuits executed in the Qutio architecture using fake\_provider in qiskit shows the same fluctuations and trends. Demonstration results on the IBMQ\_Quito computing device show that the $F_{path}$ can predict the execution fidelity of the quantum circuit effectively.

	After considering the hardware noise, the interaction path selection problem can be transformed into a search for the path with the highest metric $F_{path}$ among all paths between two qubits. This problem can be solved by the Dijkstras algorithm. The two qubits of the $set0$ or $set1$ action are used as the start and finish qubits in the interaction path selection strategy based on the Dijkstras algorithm. The strategy visits adjacent qubits from the start qubit and calculates $F_{path}$, then looks for higher $F_{path}$ vertices until the end qubit is visited. When there is only one interaction path between the start qubit and the end qubit, this path is the one with the highest $F_{path}$. This path selection strategy provides an interaction path with the highest fidelity for two logical qubits that are not adjacent to each other, so that the set operation acting on this interaction path has the lowest error rate and reduces the effect of errors in the matrix elimination process.
	
	\begin{example}
		Suppose a circuit is mapped to the coupling graph in Fig.~\ref{fig2}(b), and the two qubits mapped to $Q_7$ and $Q_9$ need to be eliminated, where the qubit mapped to $Q_7$ is the main diagonal element and the qubit mapped to $Q_9$ is 1 in the matrix. So $Q_7$ is the root node of the Steiner tree, and $Q_9$ is the child node. Since $Q_7$ and $Q_9$ are not adjacent, the $set0$ operation cannot be directly performed. Therefore, it is necessary to gradually apply $set1$ to all qubits on the path from $Q_7$ to $Q_9$. At this point, there are two interactive paths available, $path_1$ is (7-4-1-2-3-5-8-9), and $path_2$ is (7-10-12-13-14-11-8-9). The two interaction paths are shown in Fig.~\ref{Gpath}. To determine the optimal path, we calculate $F_{path_1}$ and $F_{path_2}$ respectively, and find that $F_{path_2}$ is greater than $F_{path_1}$. Therefore, $path_2$ is selected as the interactive path. 
	\end{example}
	
	\begin{figure}
		\centering
		\begin{adjustbox}{width=0.45\textwidth}
			\begin{tikzpicture}[>=stealth]
				\node at (0,0) [circle,draw=black!100,inner sep=2.2mm] {\Large9};
				\node at (0,3) [circle,draw=black!100,inner sep=2.2mm] {\Large8};
				\node at (3,3) [circle,draw=black!100,inner sep=1.5mm] {\Large11};
				\node at (6,3) [circle,draw=black!100,inner sep=1.5mm] {\Large14};
				\node at (-3,3) [circle,draw=black!100,inner sep=2.2mm] {\Large5};
				\node at (-6,3) [circle,draw=black!100,inner sep=2.2mm] {\Large3};
				\node at (-6,6) [circle,draw=black!100,inner sep=2.2mm] {\Large2};
				\node at (-6,9) [circle,draw=black!100,inner sep=2.2mm] {\Large1};
				\node at (-9,9) [circle,draw=black!100,inner sep=2.2mm] {\Large0};
				\node at (-3,9) [circle,draw=black!100,inner sep=2.2mm] {\Large4};
				\node at (0,9) [circle,draw=black!100,inner sep=2.2mm] {\Large7};
				\node at (3,9) [circle,draw=black!100,inner sep=1.5mm] {\Large10};
				\node at (6,9) [circle,draw=black!100,inner sep=1.5mm] {\Large12};
				\node at (6,6) [circle,draw=black!100,inner sep=1.5mm] {\Large13};
				\node at (9,9) [circle,draw=black!100,inner sep=1.5mm] {\Large15};
				\node at (0,12) [circle,draw=black!100,inner sep=2.2mm] {\Large6};
				\draw [thick] (0.5,3) -- (2.5,3);
				\draw [thick] (3.5,3) -- (5.5,3);
				\draw [thick] (-0.5,3) -- (-2.5,3);
				\draw [thick] (-3.5,3) -- (-5.5,3);
				\draw [thick] (0.5,9) -- (2.5,9);
				\draw [thick] (3.5,9) -- (5.5,9);
				\draw [thick] (6.5,9) -- (8.5,9);
				\draw [thick] (-0.5,9) -- (-2.5,9);
				\draw [thick] (-3.5,9) -- (-5.5,9);
				\draw [thick] (-6.5,9) -- (-8.5,9);
				\draw [thick] (0,0.5) -- (0,2.5);
				\draw [thick] (0,9.5) -- (0,11.5);
				\draw [thick] (6,3.5) -- (6,5.5);
				\draw [thick] (6,6.5) -- (6,8.5);
				\draw [thick] (-6,3.5) -- (-6,5.5);
				\draw [thick] (-6,6.5) -- (-6,8.5);
				\draw (0.8,1.5) node {\large 1.045e-2};
				\draw (6.8,4.5) node {\large 8.800e-3};
				\draw (6.8,7.5) node {\large 6.825e-3};
				\draw (0.8,10.5) node {\large 1.073e-2};
				\draw (-5.2,4.5) node {\large 1.332e-2};
				\draw (-5.2,7.5) node {\large 1.208e-2};
				\draw (1.5,3.25) node {\large 9.076e-3};
				\draw (4.5,3.25) node {\large 7.613e-3};
				\draw (1.5,9.25) node {\large 1.523e-2};
				\draw (4.5,9.25) node {\large 1.326e-2};
				\draw (7.5,9.25) node {\large 5.464e-3};
				\draw (-1.5,3.25) node {\large 7.481e-3};
				\draw (-4.5,3.25) node {\large 1.187e-2};
				\draw (-1.5,9.25) node {\large 2.458e-2};
				\draw (-4.5,9.25) node {\large 8.158e-3};
				\draw (-7.5,9.25) node {\large 1.206e-2};
				\draw [->, line width = 10pt, color=red!80, opacity = 0.3] (0,9) -- (-6,9) -- (-6,3) -- (-0.3,3) -- (-0.3,0);
				\draw [->, line width = 10pt, color=blue!80, opacity = 0.3] (0,9) -- (6,9) -- (6,3) -- (0.3,3) -- (0.3,0);
				\node at (-8,6)  {\Huge $Path_1$};
				\node at (8,6)  {\Huge $Path_2$};
			\end{tikzpicture}
		\end{adjustbox}
		\caption{There are two interaction paths between $Q_7$ and $Q_9$ in the IBMQ\_Guadalupe. The simulated fidelity of the two interacting paths is calculated by Eq.~\eqref{eq666}, and it is found that the fidelity of $F_{path_2}$ is higher, so the fidelity of the CNOT gates acting on $F_{path_2}$ is higher than those CNOT gates on $F_{path_1}$. At this point, $F_{path_2}$ is the better elimination path.}
		\label{Gpath}
	\end{figure}
	
	A minimum spanning tree for elimination is built by using the Steiner tree method, with the primary diagonal element serving as the root node. Vertices with a value of 1 in the current row or column are processed in turn, and the least weighted interaction path between these vertices and the tree is identified. All vertices on the path are then added to the tree until it contains all vertices with a value of 1. The minimum spanning tree obtained based on the above-mentioned method is referred to as a minimum-noise Steiner tree (MNST). Once minimum noise Steiner trees have been constructed for each row or column of the matrix, the matrix can be eliminated. The key-qubit priority initial mapping algorithm ensures the connectivity between the qubits during the matrix transformation. The target row matching algorithm solves the problem of finding the set of auxiliary rows of the elimination elements when processing row elimination elements. The specific flow of the NN synthesis algorithm for CNOT circuits on the general architecture is shown in Algorithm~\ref{algo4}.
	
	\begin{algorithm}[]
		\SetKwComment{Comment}{start}{end}
		\KwIn{The boolean matrix $M$, the qubit coupling diagram $G=(V,E)(\lvert V \rvert =n)$}
		\KwOut{The CNOT gate sequence applied in row operation}
		\Begin{
			$CNOT\_list \leftarrow [\ ]$;  \\
			$\pi \leftarrow KQPIMO(G,n)$;\\
			$V\_list \leftarrow \pi$;\\
			\ForEach{$i$ \rm in $V\_list$}{
				\tcp{Minimum noise Steiner tree implemented by Dijkstra}
				$T \leftarrow MNST(G,{j\lvert M_{ji}=1})$;\\
				$col\_post\_list \leftarrow $ postorder traverse $T$ from $i$;\\
				\tcp{Set the points in $T$ with a value of 0 to 1}
				\ForEach{$c$ \rm in $col\_post\_list$}{
					$k \leftarrow c.parent$;\\
					\If{$M_{ki}=0$ and $M_{ci}=1$ }{
						$M_k \leftarrow M_k\oplus M_c$;\\
						$CNOT\_list.append([c,k])$;\\
					}
				}
				\tcp{Set points in $T$ other than the root node to 0}
				\ForEach{$c$ \rm in $col\_post\_list$}{
					$l \leftarrow c.children$;\\
					$M_l \leftarrow M_l\oplus M_c$;\\
					$CNOT\_list.append([c,l])$;\\
				}
				$S_j \leftarrow TARM(M,i)$;\\	
				$T^{'} \leftarrow S_j \cup {i}$;\\
				$row\_pre\_list \leftarrow$	preorder traverse $T$ from $i$;\\
				\tcp{Set the points in $T^{'}$ with a value of 0 to 1}
				\ForEach{$r$ \rm in $row\_pre\_list$}{
					\If{$r \notin S_j$}{
						$k \leftarrow r.parent$;\\
						$M_k \leftarrow M_k\oplus M_r$;\\
						$CNOT\_list.append([r,k])$;\\
					}	    	
				}
				$row\_post\_list \leftarrow$ postorder traverse $T$ from $i$;\\
				\tcp{Set points in $T^{'}$ other than the root node to 0}
				\ForEach{$r$ \rm in $row\_post\_list$}{
					$k \leftarrow r.parent$;\\
					$M_k \leftarrow M_k\oplus M_r$;\\
					$CNOT\_list.append([r,k])$;\\
				}
				\tcp{remove qubit i from G}
				$G \leftarrow G/{i}$;\\
			}
			\Return $CNOT\_list$;
		}
		\caption{Nearest Neighbor Synthesis Algorithm for CNOT Quantum Circuits Based on Layer Convergence (LCNNS)}
		\label{algo4}
	\end{algorithm}
	
	The algorithm processes the row and column of the current main diagonal element according to the layer convergence direction. First, the algorithm generates the initial mapping method $\pi$ by algorithm~\ref{algo3} (line 3). When processing column elements, the algorithm generates a minimum noise Steiner tree based on the elements with a value of 1 in the current column (line 6). If a Steiner point exists, the child node is used to set the Steiner point to 1 (lines 8-14), and then the algorithm performs a post-order traversal to remove the main diagonal elements by $Set0$ (lines 15-19). When processing row elements, the algorithm employs the target auxiliary row matching algorithm to find the row set used to assist elimination (line 20). Then a Steiner tree is generated based on the current row and the auxiliary row set (line 21). If a Steiner point exists, the algorithm uses the child node to set the Steiner point to 1 by $Set1$ (lines 23-29). And finally, the algorithm sets the elements of the current row except the main diagonal to 0 by traversing the Steiner tree in post order (lines 31-35). After processing the row and column corresponding to the current qubit, the algorithm removes the current qubit and the edges connected to it from the coupling graph (line 36). Throughout the elimination process, the CNOT gate corresponding to each row and column operation is recorded.
	
	The following is the complexity analysis, let $n$ be the number of qubits in the coupling graph. The Steiner tree is generated by the Dijkstra algorithm (line 5), and the time complexity is $O(n^2)$. The time complexity of traversing the Steiner tree (line 6) is $O(n^2)$. The operations in the first foreach loop (lines 7-11) have constant time complexity, so the time complexity of the foreach loop is $Q(n)$. Similarly, the time complexity of the foreach loop (line 14) is also $Q(n)$. The time complexity of the TARM algorithm (line 19) is $Q(n \cdot n!)$. A new tree is constructed by using the Prim algorithm (line 20), and the time complexity is $O(n^2)$. The time complexity of obtaining the preorder traversal list (line 21) is $Q(n)$. The time complexity of the two foreach loops (lines 22 and 30) are both $Q(n)$. The time complexity of traversing the list (line 28) is $O(n^2)$. So the total time complexity of algorithm~\ref{algo4} is shown in Eq.~\eqref{eq8}.
	\begin{eqnarray}
		O(n(n^2+n^2+n+n!n+n^2+n+n^2+n))\nonumber\\
		=O(4n^3+n^2 n!+4n)
		\label{eq8}
	\end{eqnarray}
	where $n$ represents the number of qubits in the coupling graph. In summary, the time complexity of Algorithm~\ref{algo4} is dominated by the time complexity of the TARM algorithm. The complexity is further simplified to $O(n^2 n!)$.
	
	\subsection{Algorithm example}
\begin{figure}[]
		\subfigure[]{
			\raisebox{-1.2cm}{
				\begin{adjustbox}{width=0.2\textwidth}
					\includegraphics{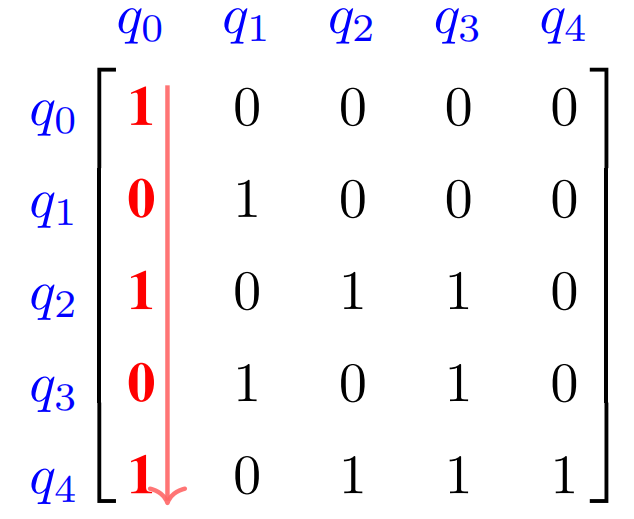}
				\end{adjustbox}
			}
			\hspace{0.1cm}
			\begin{adjustbox}{width=0.25\textwidth}
				
				\begin{tikzpicture}
					[>=stealth,baseline=-3cm,node distance=1.9cm, thick, inner sep=0pt,minimum size=8mm,
					black/.style={circle,draw=black!100},
					red/.style={circle,draw=black!100,fill=red!20},
					green/.style={circle,draw=black!100,fill=green!20},
					dashgray/.style={circle,dashed,draw=black!100,fill=gray!10},
					dashgreen/.style={circle,dashed,draw=black!100,fill=green!20}]
					\node[dashgreen]		(Q3)					             {$q_3$};
					\node[green]		         (Q2)  [below=of Q3]   {$q_2$};
					\node[black]		          (Q1)  [below=of Q2]  {$q_1$};
					\node[red]                     (Q0) [left=of Q3]        {$q_0$};
					\node[green] 	             (Q4) [right=of Q3]     {$q_4$};
					\path []
					(Q0)   edge		(Q3)
					edge [->,bend right,color=blue!50]		(Q3)
					(Q3)   edge      (Q4)
					edge [->,bend left=45,color=blue!50]		(Q4)
					edge      (Q2)
					edge [->,bend right=45,color=blue!50]		(Q2)
					(Q2)   edge      (Q1)
					(Q4)   edge	 [->,bend left,color=blue!50]		(Q3);
					
					\draw (Q0) to node [above] {$2.523e^{-2}$} (Q3);
					\draw (Q3) to node [above] {$8.710e^{-3}$} (Q4);
					\draw (Q2) to node [sloped,above] {$1.851e^{-2}$} (Q3);
					\draw (Q1) to node [sloped,above] {$1.595e^{-2}$} (Q2);
					\draw (1.4,1) node {\large{\textcircled{\scriptsize{2}}}\large};
					\draw (1.4,-0.85) node {\large{\textcircled{\scriptsize{1}}}\large};
					\draw (-1,-1.4) node {\large{\textcircled{\scriptsize{3}}}\large};
					\draw (-1.4,-0.85) node {\large{\textcircled{\scriptsize{4}}}\large};
				\end{tikzpicture}
			\end{adjustbox}
		}
		\subfigure[]{
			\raisebox{-1.2cm}{
				\begin{adjustbox}{width=0.2\textwidth}
					\includegraphics{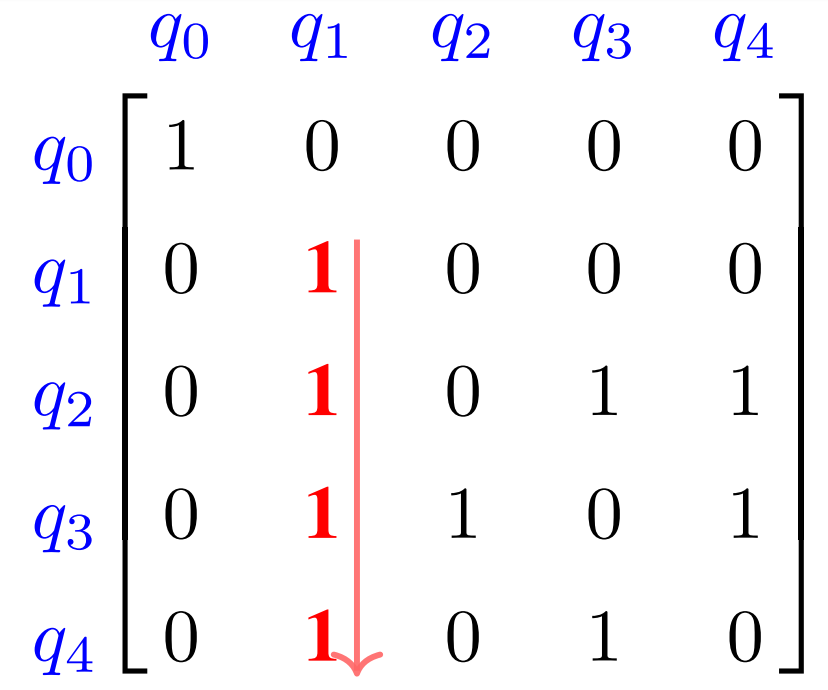}
				\end{adjustbox}
			}
			\hspace{0.1cm}
			\begin{adjustbox}{width=0.25\textwidth}
				\begin{tikzpicture}
					[>=stealth,baseline=-3cm,node distance=1.9cm, thick, inner sep=0pt,minimum size=8mm,
					black/.style={circle,draw=black!100},
					red/.style={circle,draw=black!100,fill=red!20},
					green/.style={circle,draw=black!100,fill=green!20},
					dashgray/.style={circle,dashed,draw=gray!100,fill=gray!10},
					dashgreen/.style={circle,dashed,draw=black!100,fill=green!10}]
					\node[green]		(Q3)					             {$q_3$};
					\node[green]		         (Q2)  [below=of Q3]   {$q_2$};
					\node[red]		          (Q1)  [below=of Q2]  {$q_1$};
					\node[dashgray]                     (Q0) [left=of Q3]        {$q_0$};
					\node[green] 	             (Q4) [right=of Q3]     {$q_4$};
					\path []
					(Q0)   edge[dashed,draw=gray]		(Q3)
					(Q3)   edge      (Q4)
					edge [->,bend left=45,color=blue!50]		(Q4)
					edge      (Q2)
					(Q2)	edge [->,bend left=45,color=blue!50]		(Q3)
					(Q2)   edge      (Q1)
					(Q1)   edge	 [->,bend left=45,color=blue!50]		(Q2);
					
					\draw (Q3) to node [above] {$8.710e^{-3}$} (Q4);
					\draw (Q2) to node [sloped,above] {$1.851e^{-2}$} (Q3);
					\draw (Q1) to node [sloped,above] {$1.595e^{-2}$} (Q2);
					\draw (1.4,1) node {\large{\textcircled{\scriptsize{1}}}\large};
					\draw (-1,-1.5) node {\large{\textcircled{\scriptsize{2}}}\large};
					\draw (-1,-4.2) node {\large{\textcircled{\scriptsize{3}}}\large};
				\end{tikzpicture}
			\end{adjustbox}
		}
		\subfigure[]{
			\raisebox{-1.2cm}{
				\begin{adjustbox}{width=0.2\textwidth}
					\includegraphics{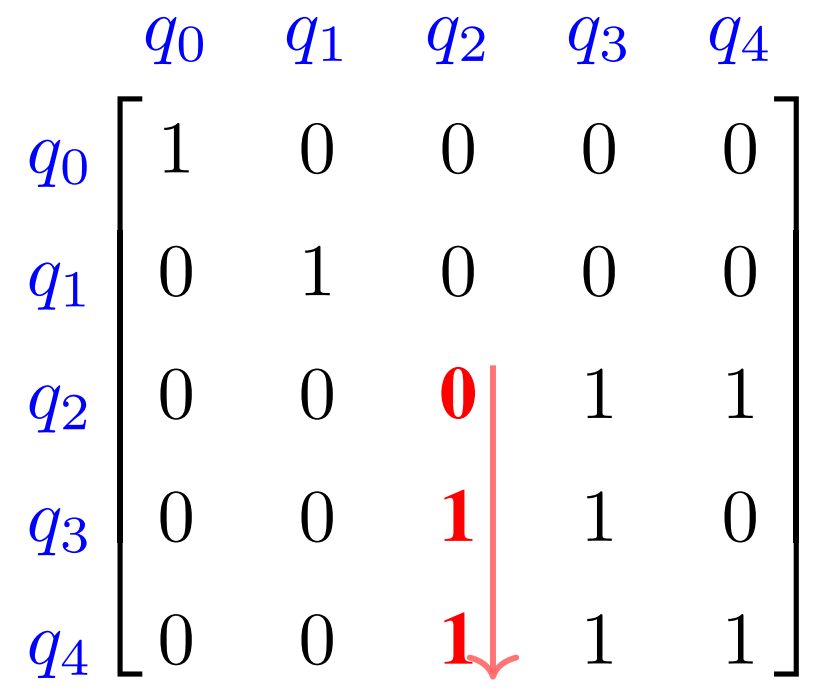}
				\end{adjustbox}
			}
			\hspace{0.1cm}
			\begin{adjustbox}{width=0.25\textwidth}
				\begin{tikzpicture}
					[>=stealth,baseline=-3cm,node distance=1.9cm, thick, inner sep=0pt,minimum size=8mm,
					black/.style={circle,draw=black!100},
					red/.style={circle,draw=black!100,fill=red!20},
					green/.style={circle,draw=black!100,fill=green!20},
					dashgray/.style={circle,dashed,draw=gray!100,fill=gray!10},
					dashgreen/.style={circle,dashed,draw=black!100,fill=green!10}]
					\node[green]		(Q3)					             {$q_3$};
					\node[red,dashed]		         (Q2)  [below=of Q3]   {$q_2$};
					\node[dashgray]		          (Q1)  [below=of Q2]  {$q_1$};
					\node[dashgray]                     (Q0) [left=of Q3]        {$q_0$};
					\node[green] 	             (Q4) [right=of Q3]     {$q_4$};
					\path []
					(Q0)   edge[dashed,draw=gray]		(Q3)
					(Q3)   edge      (Q4)
					edge [->,bend left=45,color=blue!50]		(Q4)
					edge      (Q2)
					edge [->,bend left,color=blue!50]		(Q2)
					(Q2)	edge [->,bend left=45,color=blue!50]		(Q3)
					(Q2)   edge[dashed,draw=gray]      (Q1);
					\draw (Q3) to node [above] {$8.710e^{-3}$} (Q4);
					\draw (Q2) to node [sloped,above] {$1.851e^{-2}$} (Q3);
					\draw (1.4,1) node {\large{\textcircled{\scriptsize{2}}}\large};
					\draw (-1,-1.5) node {\large{\textcircled{\scriptsize{3}}}\large};
					\draw (0.85,-1.5) node {\large{\textcircled{\scriptsize{1}}}\large};
				\end{tikzpicture}
			\end{adjustbox}
		}
		\subfigure[]{
			\raisebox{-1.2cm}{
				\begin{adjustbox}{width=0.2\textwidth}
					\includegraphics{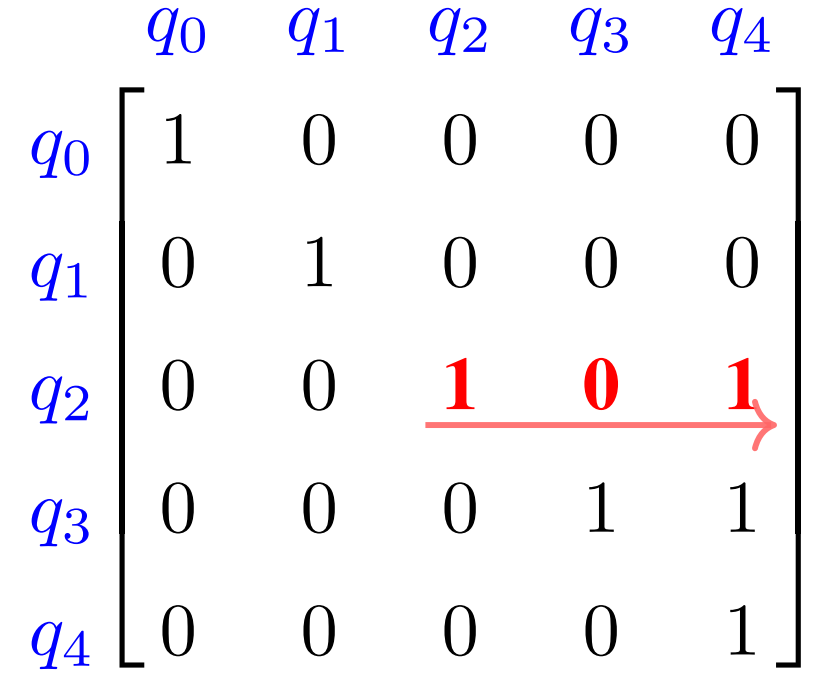}
				\end{adjustbox}
			}
			\hspace{0.1cm}
			\begin{adjustbox}{width=0.25\textwidth}
				\begin{tikzpicture}
					[>=stealth,baseline=-3cm,node distance=1.9cm, thick, inner sep=0pt,minimum size=8mm,
					black/.style={circle,draw=black!100},
					red/.style={circle,draw=black!100,fill=red!20},
					blue/.style={circle,draw=black!100,fill=blue!20},
					green/.style={circle,draw=black!100,fill=green!20},
					dashgray/.style={circle,dashed,draw=gray!100,fill=gray!10},
					dashgreen/.style={circle,dashed,draw=black!100,fill=green!10}]
					\node[blue,dashed]		(Q3)					             {$q_3$};
					\node[red]		         (Q2)  [below=of Q3]   {$q_2$};
					\node[dashgray]		          (Q1)  [below=of Q2]  {$q_1$};
					\node[dashgray]                     (Q0) [left=of Q3]        {$q_0$};
					\node[green] 	             (Q4) [right=of Q3]     {$q_4$};
					\path []
					(Q0)   edge[dashed,draw=gray]		(Q3)
					(Q3)   edge      (Q4)
					edge [->,bend left,color=blue!50]		(Q2)
					edge [->,bend right=45,color=blue!50]		(Q2)
					edge      (Q2)
					(Q2)   edge[dashed,draw=gray]      (Q1)
					(Q4)    edge [->,bend right=45,color=blue!50]     (Q3);
					\draw (Q3) to node [above] {$8.710e^{-3}$} (Q4);
					\draw (Q2) to node [sloped,above] {$1.851e^{-2}$} (Q3);
					\draw (1.4,1) node {\large{\textcircled{\scriptsize{2}}}\large};
					\draw (-1,-1.5) node {\large{\textcircled{\scriptsize{3}}}\large};
					\draw (0.85,-1.5) node {\large{\textcircled{\scriptsize{1}}}\large};
				\end{tikzpicture}
			\end{adjustbox}
		}
		\caption{Example of synthesis process for the circuit in Fig.~\ref{fig3}(a).}
		\label{fig9}      
	\end{figure}

	To illustrate the NN synthesis process, we consider the CNOT circuit shown in Fig.~\ref{fig3}(a) and use the IBMQ\_Quito architecture as an example. The synthesis process is presented in Fig.~\ref{fig9}. According to the key-qubit priority initial mapping optimization algorithm, the initial mapping \{$q_0 \rightarrow Q_0, q_1 \rightarrow Q_4, q_2 \rightarrow Q_3,q_3 \rightarrow Q_1,q_4 \rightarrow Q_2$\} that satisfies the key-qubit priority model is obtained.

	First, the algorithm processes the first row of the first column. The first column of the 1 element corresponds to the qubits $q_0$, $q_2$, $q_4$. The Steiner tree is generated in the coupling graph, and the algorithm performs post-order traversal of the current Steiner tree. If the current node value is 1 and the parent node value is 0, the current node is used to set its parent node to 1. So $q_4$ is used to set $q_3$ to 1 first, and CNOT(4,3) is executed. Next, the entire Steiner tree is traversed posteriorly, and $q_4$, $q_2$, and $q_3$ are traversed in order to apply the parent node of the current node to the current row, i.e., CNOT(3,4), CNOT(3,2), and CNOT(0,3) are applied. Because the first row is already a unit vector, there is no need to process the first row.
	
	Next, the second column and the second row are processed. The qubits corresponding to the 1 element in the second column are $q_1$, $q_2$, $q_3$, $q_4$. The Steiner tree is generated in the coupling graph, and the current Steiner tree is traversed in the post-order. There are no Steiner points in the current Steiner tree, so no operation is needed. Then, the algorithm post-order traverses the whole Steiner tree, traverses $q_4$,  $q_3$,  $q_2$ in turn, and applies the row where the parent node of the current node is located to the current row, i.e., applies CNOT(3,4), CNOT(3,2), CNOT(1,2). Since the second row is already a unit vector, there is no need to process the second row.
	
	Finally, the third column and the third row are processed. The qubits corresponding to the 1 element in the third column are $q_3$ and $q_4$. The Steiner tree is generated in the coupling graph, and the current Steiner tree is traversed in the post-order. The qubit $q_3$ is used to set $q_2$ to 1 first, and then apply CNOT(3,2). Then, the algorithm traverses the whole Steiner tree in post-order, traverses $q_4$ and $q_3$ in turn, and applies the row where the parent node of the current node is located to the current row, that is, applies CNOT(3,4), CNOT(3,2). At this point, the third row is not a unit vector and needs to be processed. By using the target-aided rows matching algorithm, the third row can be obtained from the fifth row and the unit matrix $e_i$=[0,0,1,0,0]. So the Steiner tree is generated based on $q_2$, $q_4$. After the first preorder traversal of the Steiner tree, because $q_3$ is not in the set of target rows, it is necessary to apply the corresponding row of $q_3$ to the corresponding row of $q_4$, i.e., apply CNOT(3,4). Then the algorithm post-order traverses the Steiner tree, traverses $q_4$ and $q_3$ in turn, and applies the row where the current node is located to the row where the parent node is located, i.e., apply CNOT(4,3), CNOT(3,2). At this point, the whole matrix becomes a unit matrix, and the NN synthesis ends.

	\section{Demonstration results and analysis}
	\label{sec:experimental}
	In this section, the performance of the proposed synthesis method is evaluated using various metrics. The configurations used for the evaluations are first provided, followed by a detailed presentation of results for synthesis algorithm across various benchmarks. A specific analysis of these results is then conducted to highlight the effectiveness of the synthesis method.
	
	\subsection{Demonstration configurations}
	\begin{enumerate}[label=\arabic{enumi})]
		 \item	Runtime Environment: The algorithms are all implemented in Python language, and the code environment is the macOS Big Sur (11.2.3) operating system with Apple M1 Pro octa-core processing and 16GB RAM.
		 \item Quantum Computing Resources: The quantum cloud platforms used for the demonstration include OriginQ~\cite{origin}, Quafu~\cite{quafu} and IBMQ~\cite{IBMQ}. The coupling graphs and detailed parameters of the quantum computing devices on these platforms are shown in Appendices~\ref{Coupling} and~\ref{Parameter}.
		 \item Benchmarks: First, the 16- and 20-qubit random CNOT circuits given in \cite{kissinger2020cnot} are demonstrated on the QX5 and Tokyo fake\_provider simulators, respectively, with CNOT gate sizes of 4, 8, 16, 32, 64, 128 and 256.
         Secondly, a selection of randomly generated CNOT circuits including 5 and 16 qubits, with gate levels ranging from 10 to 10,000, is chosen for demonstrated on the Quito and Guadalupe fake\_provider simulators, respectively. Finally, the synthesis of the Bernstein-Vazirani algorithm on different platforms with different architectures is utilized to illustrate the generality of the proposed method.
		 \item Metrics: CONT, the number of CNOT gates following synthesis; Depth, the depth of the circuit following synthesis; and Fidelity, the circuit fidelity following synthesis. For a circuit consisting of pure CNOT gates, the calculation of fidelity is defined as the ratio of the number of simulations where the output states remain all zeros to the total number of simulations. This fidelity is calculated by Eq.~\eqref{eq_fidelity},
		 \begin{equation}
		 	F=\frac{N_{\text {zero }}}{N_{\text {total }}}
		 	\label{eq_fidelity}
		 \end{equation}
		 where $N_{\text{zero}}$ represents the number of simulations in which the output states remain all zeros, and $N_{\text{total}}$ represents the total number of simulations. This approach is suitable for studies using CNOT circuits, as the ideal output for all-zero input states is also all zeros, making this ratio an effective measure of the circuit's fidelity under noisy conditions. It is important to note that this definition of fidelity is specifically the ratio of successful all-zero output states to the total number of simulations, and should not be confused with other variants like estimated success probability (ESP) or total variance distance (TVD). For circuits containing single-qubit gates, ESP or TVD are better ways to calculate fidelity.
    \end{enumerate}

\subsection{Demonstration results}

\subsubsection{Comparison with synthesis}
In this part, the performance of the proposed LCNNS method is evaluated by comparing it with the synthesis approaches presented in Refs. \cite{kissinger2020cnot}, \cite{zhu2022physical}, and \cite{wu2023optimization}. The comparison is carried out on the QX5 and Tokyo fake\_provider simulators, respectively. The CNOT circuits, as introduced in the first part of the benchmarks, are utilized for this evaluation. For each circuit size, 20 instances are considered, and the results are averaged across these instances. In addition, circuit fidelity is performed by the fake\_provider simulator~\cite{qiskit}, which is part of the Qiskit development package provided by the IBMQ cloud platform. The fake\_provider simulator contains important information about the quantum system, such as coupling map, basis gates, qubit properties (T1, T2, error rate, etc.) which are essential for performing noisy simulation of the system. 

\begin{figure*}[]
   	\subfigure[QX5]{
   		\begin{adjustbox}{width=0.72\textwidth}
   			\includegraphics{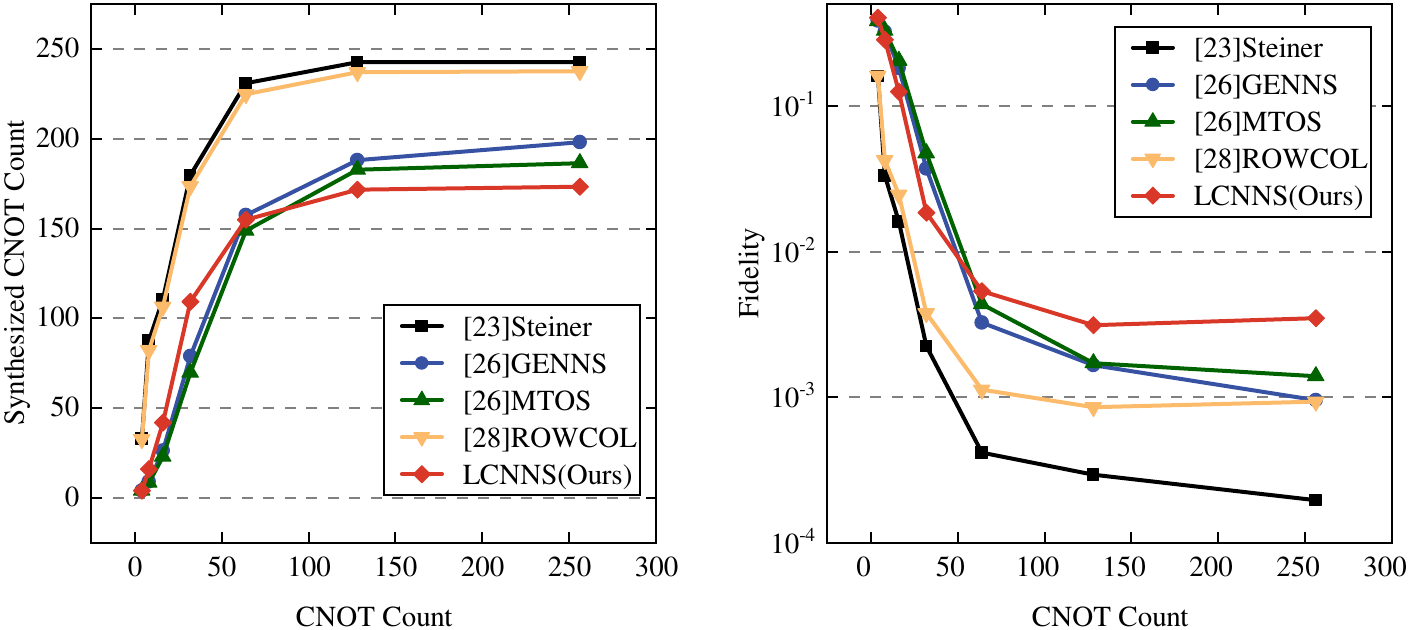}
   		\end{adjustbox}
   	}
   	\subfigure[Tokyo]{
   		\begin{adjustbox}{width=0.72\textwidth}
   			\includegraphics{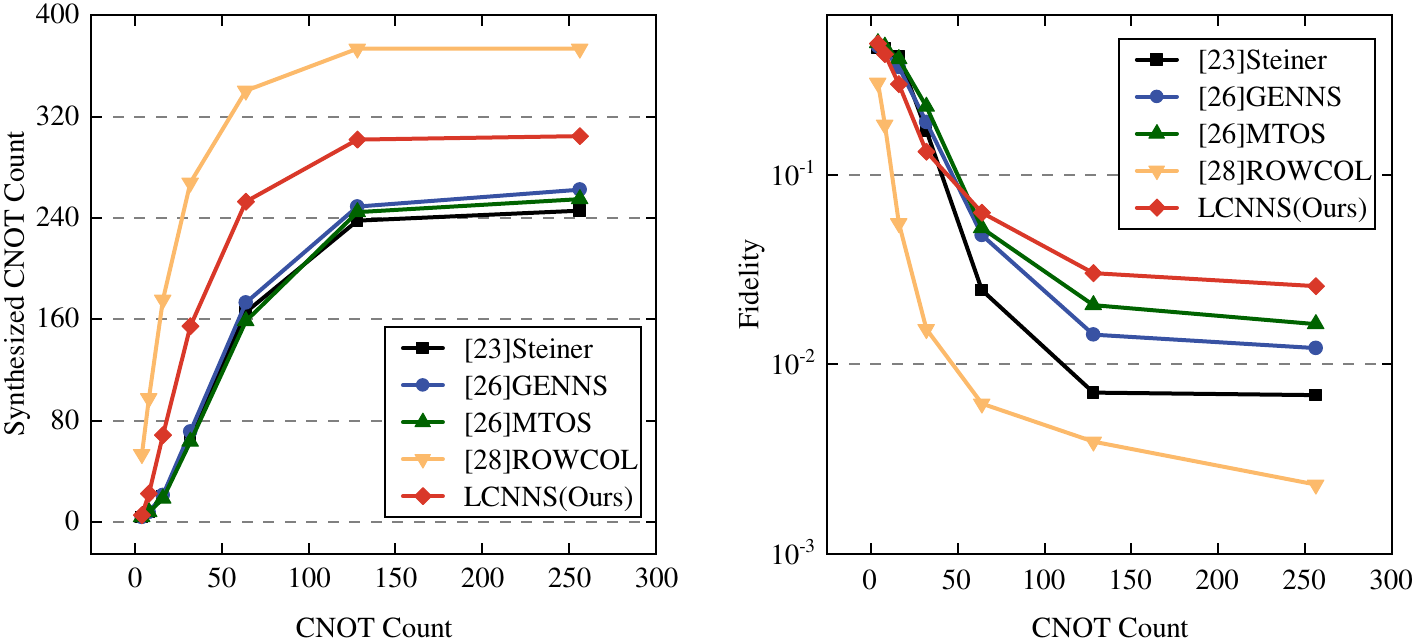}
   		\end{adjustbox}
   	}
   	\caption{Comparison of synthesized CNOT count and fidetity on IBMQ QX5 and Tokyo.}
   	\label{figex1}
   \end{figure*}

 The results, shown in Fig. \ref{figex1}, indicate that on the QX5 architecture, as the number of initial CNOT gates increases, the number of synthesized CNOT gates is lower than that of the comparison methods, and the fidelity is also higher. This trend becomes more evident as the number of CNOT gates increases. Specifically, on the QX5 architecture, the LCNNS method achieves a 21.3\% improvement in fidelity over the other methods. On the Tokyo architecture, the methods proposed in \cite{kissinger2020cnot} and \cite{zhu2022physical} are more suitable for architectures that include Hamiltonian paths. In contrast, the methods in  \cite{wu2023optimization} and the LCNNS sacrifice the advantage in gate count for the generality of the architecture. Therefore, the synthesized CNOT gate count in  \cite{wu2023optimization} and the LCNNS method is higher than that of other comparison methods. Despite this increased gate count, the noise-aware LCNNS method maintains superior fidelity, achieving an average improvement of 7.0\% over the best competing methods. The overall average optimization rate across both the QX5 and Tokyo architectures is 14.2\%. This result demonstrates that the LCNNS method can effectively optimize circuits on both architectures, particularly showing advantages in maintaining quantum state fidelity.

\subsubsection{Comparison with compilers}
\begin{table*}[]
		\caption{The test data of 5-qubit random CNOT circuits on IBMQ\_Quito's fake\_provider simulator in Qiskit provided by the IBMQ cloud platform.}
		\setlength{\tabcolsep}{1.1pt} 
		\renewcommand{\arraystretch}{1.5} 
		\begin{ruledtabular}
\begin{tabular}{cccccccccccccc}
\multirow{2}{*}{Circuit\footnote{The number 5 indicates 5 qubits, and the following number represents the number of CNOT gates, ranging from 10 to 10000.}}      & \multicolumn{3}{c}{Qiskit}                  & \multicolumn{3}{c}{Tket}                    & \multicolumn{3}{c}{HA}                      & \multicolumn{3}{c}{LCNNS}                  & \multirow{2}{*}{Imp\footnote{Imp: the Imp is calculated as the ratio of the fidelity achieved using LCNNS to the highest fidelity among other methods. A ratio greater than 1 indicates optimization, with a larger ratio signifying more significant improvement.}} \\ \cline{2-13}
                              & CNOT\footnotemark[3]  & Depth\footnotemark[3] & Fidelity\footnotemark[3]                        & CNOT  & Depth & Fidelity                         & CNOT  & Depth & Fidelity                         & CNOT & Depth & Fidelity                         &                      \\ \hline
\multicolumn{1}{c|}{5\_10}    & 19    & 16    & \multicolumn{1}{c|}{0.6045} & 19    & 23    & \multicolumn{1}{c|}{0.6025} & 23    & 21    & \multicolumn{1}{c|}{0.5586} & 14   & 15    & \multicolumn{1}{c|}{0.6670} & 1.10                 \\
\multicolumn{1}{c|}{5\_20}    & 35    & 26    & \multicolumn{1}{c|}{0.4971} & 37    & 37    & \multicolumn{1}{c|}{0.3584} & 46    & 39    & \multicolumn{1}{c|}{0.3486} & 14   & 14    & \multicolumn{1}{c|}{0.6543} & 1.32                 \\
\multicolumn{1}{c|}{5\_30}    & 69    & 61    & \multicolumn{1}{c|}{0.1914} & 62    & 55    & \multicolumn{1}{c|}{0.2217} & 68    & 61    & \multicolumn{1}{c|}{0.2383} & 16   & 15    & \multicolumn{1}{c|}{0.6162} & 2.59                 \\
\multicolumn{1}{c|}{5\_40}    & 75    & 66    & \multicolumn{1}{c|}{0.2217} & 72    & 86    & \multicolumn{1}{c|}{0.2129} & 100   & 83    & \multicolumn{1}{c|}{0.1582} & 12   & 13    & \multicolumn{1}{c|}{0.6855} & 3.09                 \\
\multicolumn{1}{c|}{5\_50}    & 115   & 97    & \multicolumn{1}{c|}{0.1113} & 103   & 110   & \multicolumn{1}{c|}{0.1152} & 110   & 89    & \multicolumn{1}{c|}{0.1406} & 16   & 16    & \multicolumn{1}{c|}{0.6357} & 4.52                 \\
\multicolumn{1}{c|}{5\_100}   & 225   & 196   & \multicolumn{1}{c|}{0.0596} & 195   & 188   & \multicolumn{1}{c|}{0.0557} & 209   & 176   & \multicolumn{1}{c|}{0.0645} & 25   & 24    & \multicolumn{1}{c|}{0.5049} & 7.83                 \\
\multicolumn{1}{c|}{5\_200}   & 492   & 441   & \multicolumn{1}{c|}{0.0635} & 430   & 444   & \multicolumn{1}{c|}{0.0625} & 482   & 404   & \multicolumn{1}{c|}{0.0723} & 21   & 20    & \multicolumn{1}{c|}{0.5205} & 7.20                 \\
\multicolumn{1}{c|}{5\_500}   & 1072  & 966   & \multicolumn{1}{c|}{0.0625} & 977   & 986   & \multicolumn{1}{c|}{0.0605} & 1094  & 919   & \multicolumn{1}{c|}{0.0566} & 11   & 11    & \multicolumn{1}{c|}{0.7197} & 11.52                \\
\multicolumn{1}{c|}{5\_1000}  & 2296  & 2041  & \multicolumn{1}{c|}{0.0723} & 2028  & 2123  & \multicolumn{1}{c|}{0.0635} & 2436  & 2023  & \multicolumn{1}{c|}{0.0547} & 28   & 26    & \multicolumn{1}{c|}{0.4717} & 6.53                 \\
\multicolumn{1}{c|}{5\_2000}  & 4565  & 4033  & \multicolumn{1}{c|}{0.0498} & 4042  & 4177  & \multicolumn{1}{c|}{0.0576} & 4740  & 3923  & \multicolumn{1}{c|}{0.0566} & 20   & 19    & \multicolumn{1}{c|}{0.5479} & 9.51                 \\
\multicolumn{1}{c|}{5\_5000}  & 11547 & 10249 & \multicolumn{1}{c|}{0.0635} & 10029 & 10314 & \multicolumn{1}{c|}{0.0654} & 11984 & 9926  & \multicolumn{1}{c|}{0.0645} & 20   & 19    & \multicolumn{1}{c|}{0.5459} & 8.34                 \\
\multicolumn{1}{c|}{5\_10000} & 23023 & 20529 & \multicolumn{1}{c|}{0.0723} & 20047 & 20713 & \multicolumn{1}{c|}{0.0566} & 23480 & 19592 & \multicolumn{1}{c|}{0.0557} & 22   & 21    & \multicolumn{1}{c|}{0.5479} & 7.58                 \\ 
\end{tabular}
\end{ruledtabular}
\footnotetext[3]{CONT, Depth, Fidelity: the number of CNOT gates, the depth of the circuit, and the circuit fidelity following compilation using four different methods: Qiskit, Tket, HA, and LCNNS.}
\label{table1}
\end{table*}

    The second part of the benchmarks is implemented on the Quito and Guadalupe fake\_provider simulators, respectively, to demonstrate the comparison between the LCNNS method and mainstream compilers. The test data for this benchmark circuit is presented in Table~\ref{table1} and Table~\ref{table2}. And the corresponding data comparison graph is shown in Figs.~\ref{fig12}-\ref{fig13}. It is worth noting that four different methods are compared for synthesizing the CNOT circuit, namely Qiskit, Tket, HA and LCNNS. Qiskit is a quantum computing framework developed by IBM that can be used to build and simulate quantum circuits. Tket is a quantum compiler developed by Cambridge Quantum Computing that converts advanced quantum algorithms into quantum circuits. HA is a compilation method proposed in \cite{niu2020hardware}. In the first three compilers, the default mapping strategy is selected for comparison. In terms of fidelity, the fidelity of the circuits in Tables~\ref{table1} and ~\ref{table2} is optimized up to 5.7 times and 243.4 times on average.

    In the NISQ era of quantum computing, one significant source of noise is the CNOT gate. Thus, reducing the number of CNOT gates during the synthesis process is instrumental in enhancing fidelity. The theoretical upper bound for the number of CNOT gates in an NN synthesized CNOT circuit has been established as $2n^2$,  $n$ represents the number of qubits in the circuit, as demonstrated in~\cite{wu2023optimization}. Tables \ref{table1} and \ref{table2} show that the number of CNOT gates under the LCNNS method is consistently below this upper bound, indirectly validating the theoretical correctness. In this study, noise factors are explicitly considered during the synthesis process, which further aids in improving the fidelity of the circuits. Therefore, the optimization rates presented in Tables \ref{table1} and \ref{table2} are considered in the context of this fidelity enhancement.


\begin{table*}[]
\caption{The test data of 16-qubit random CNOT circuits on the IBMQ\_Guadalupe's fake\_provider simulator in Qiskit provided by the IBMQ cloud platform.}
		\setlength{\tabcolsep}{1.1pt} 
		\renewcommand{\arraystretch}{1.5} 
		\begin{ruledtabular}
\begin{tabular}{cccccccccccccc}
\multirow{2}{*}{Circuit}       & \multicolumn{3}{c}{Qiskit}                  & \multicolumn{3}{c}{Tket}                    & \multicolumn{3}{c}{HA}                      & \multicolumn{3}{c}{LCNNS}                  & \multirow{2}{*}{Imp} \\ \cline{2-13}
                               & CNOT  & Depth & Fidelity                         & CNOT  & Depth & Fidelity                         & CNOT  & Depth & Fidelity                         & CNOT & Depth & Fidelity                         &                      \\ \cline{0-13}
\multicolumn{1}{c|}{16\_10}    & 26    & 15    & \multicolumn{1}{c|}{0.7842} & 31    & 19    & \multicolumn{1}{c|}{0.7510} & 62    & 28    & \multicolumn{1}{c|}{0.6594} & 125  & 87    & \multicolumn{1}{c|}{0.4463} & 0.57                 \\
\multicolumn{1}{c|}{16\_20}    & 61    & 30    & \multicolumn{1}{c|}{0.6611} & 60    & 44    & \multicolumn{1}{c|}{0.6543} & 129   & 56    & \multicolumn{1}{c|}{0.5251} & 148  & 86    & \multicolumn{1}{c|}{0.4492} & 0.68                 \\
\multicolumn{1}{c|}{16\_30}    & 105   & 40    & \multicolumn{1}{c|}{0.5508} & 105   & 55    & \multicolumn{1}{c|}{0.5576} & 156   & 68    & \multicolumn{1}{c|}{0.4756} & 210  & 135   & \multicolumn{1}{c|}{0.3369} & 0.60                 \\
\multicolumn{1}{c|}{16\_40}    & 151   & 69    & \multicolumn{1}{c|}{0.3992} & 173   & 90    & \multicolumn{1}{c|}{0.4199} & 224   & 113   & \multicolumn{1}{c|}{0.3389} & 210  & 130   & \multicolumn{1}{c|}{0.3115} & 0.74                 \\
\multicolumn{1}{c|}{16\_50}    & 240   & 96    & \multicolumn{1}{c|}{0.2815} & 293   & 147   & \multicolumn{1}{c|}{0.2168} & 374   & 164   & \multicolumn{1}{c|}{0.2009} & 258  & 149   & \multicolumn{1}{c|}{0.2461} & 0.87                 \\
\multicolumn{1}{c|}{16\_100}   & 475   & 198   & \multicolumn{1}{c|}{0.1033} & 576   & 307   & \multicolumn{1}{c|}{0.0596} & 651   & 319   & \multicolumn{1}{c|}{0.0635} & 242  & 129   & \multicolumn{1}{c|}{0.2568} & 2.49                 \\
\multicolumn{1}{c|}{16\_200}   & 1043  & 425   & \multicolumn{1}{c|}{0.0061} & 1261  & 692   & \multicolumn{1}{c|}{0.0029} & 1535  & 697   & \multicolumn{1}{c|}{0.0037} & 272  & 158   & \multicolumn{1}{c|}{0.2246} & 36.80                \\
\multicolumn{1}{c|}{16\_500}   & 2751  & 1056  & \multicolumn{1}{c|}{0.0002} & 3115  & 1660  & \multicolumn{1}{c|}{0.0002} & 3733  & 1654  & \multicolumn{1}{c|}{0.0001} & 250  & 141   & \multicolumn{1}{c|}{0.2627} & 1076.00              \\
\multicolumn{1}{c|}{16\_1000}  & 5634  & 2208  & \multicolumn{1}{c|}{0.0001} & 6091  & 3172  & \multicolumn{1}{c|}{0.0010} & 7743  & 3660  & \multicolumn{1}{c|}{0.0001} & 259  & 137   & \multicolumn{1}{c|}{0.2422} & 248.00               \\
\multicolumn{1}{c|}{16\_2000}  & 11374 & 4633  & \multicolumn{1}{c|}{0.0002} & 12116 & 6452  & \multicolumn{1}{c|}{0.0005} & 15236 & 6997  & \multicolumn{1}{c|}{0.0002} & 254  & 137   & \multicolumn{1}{c|}{0.2383} & 488.00               \\
\multicolumn{1}{c|}{16\_5000}  & 29184 & 11545 & \multicolumn{1}{c|}{0.0002} & 31231 & 16807 & \multicolumn{1}{c|}{0.0002} & 38507 & 18106 & \multicolumn{1}{c|}{0.0002} & 247  & 133   & \multicolumn{1}{c|}{0.2637} & 1080.00              \\
\multicolumn{1}{c|}{16\_10000} & 58942 & 23212 & \multicolumn{1}{c|}{0.0003} & 63282 & 33848 & \multicolumn{1}{c|}{0.0003} & 77780 & 35606 & \multicolumn{1}{c|}{0.0002} & 282  & 159   & \multicolumn{1}{c|}{0.2422} & 793.60               \\ 
\end{tabular}
		\end{ruledtabular}
   \begin{tablenotes}
   \footnotesize
    \item  * The notes in Table~\ref{table2} are consistent with Table~\ref{table1}.
   \end{tablenotes}
		\label{table2}
\end{table*}

\subsubsection{Synthesis on the BV algorithm}
   
   \begin{figure}[]
   	\centering
   	\begin{adjustbox}{width=0.45\textwidth}
   		\begin{quantikz}[row sep={0.7cm,between origins}]
   			\lstick{$q_0$}  & \gate{H} \gategroup[7,steps=2,style={dashed,
   				rounded corners, fill=blue!20, inner xsep=0pt},background]{{single-qubit gates}} & \qw  & \ctrl{6} \gategroup[7,steps=6,style={dashed,
   				rounded corners, fill=red!20, inner xsep=0pt},background]{{CNOT subcircuit}}  & \qw  & \qw & \qw & \ \ldots\ \qw & \qw  & \gate{H}\gategroup[7,steps=1,style={dashed,
   				rounded corners, fill=blue!20, inner xsep=0pt},background]{{single-qubit gates}} & \meter{} & \qw\\
   			\lstick{$q_1$}  & \gate{H} & \qw  & \qw  & \ctrl{5}  & \qw & \qw  & \ \ldots\ \qw & \qw & \gate{H} & \meter{}& \qw\\
   			\lstick{$q_2$}  & \gate{H} & \qw  & \qw  & \qw & \ctrl{4}   & \qw & \ \ldots\ \qw & \qw & \gate{H}& \meter{} & \qw\\
   			\lstick{$q_3$}  & \gate{H} & \qw  & \qw  & \qw & \qw & \ctrl{3}   & \ \ldots\ \qw & \qw & \gate{H} & \meter{}& \qw\\
   			\lstick{$\vdots$}  & \wave&&&&&&&&& \rstick{$\vdots$}\\
   			\lstick{$q_{n-2}$} & \gate{H} & \qw & \qw  & \qw  & \qw & \qw  & \ \ldots\ \qw & \ctrl{1} & \gate{H}& \meter{} & \qw\\
   			\lstick{$q_{n-1}$} &  \gate{H} &  \gate{Z} & \targ{}  & \targ{}  & \targ{} & \targ{} & \ \ldots\ \qw & \targ{} & \qw & \qw 
   		\end{quantikz}
   	\end{adjustbox}   
   	\caption{Quantum circuit representation of the Bernstein-Vazirani algorithm. Inside the blue box are single-qubit gates, which are not subject to the NN constraints of the quantum architecture. In the red box are CNOT gates, and the NN of the whole circuit can be achieved by simply applying the LCNNS method to this portion of consecutive CNOT gates and then recombining them with the single-qubit gates.}
   	\label{Bernstein-Vazirani} 
   \end{figure}
   
      In order to extend the CNOT circuit synthesis method to overcome its limitation of being applicable only to CNOT circuits, the Bernstein-Vazirani quantum algorithm is used as an example to illustrate the extensibility of the CNOT circuit synthesis method. Bernstein-Vazirani is a quantum algorithm containing some single-qubit gates and CNOT gates, where the CNOT gates are continuous, as shown in Fig.~\ref{Bernstein-Vazirani}. We synthesize this part of the CNOT circuits by using the LCNNS method, after adding the single-qubit gates into the synthesized circuits, making them equivalent to the original circuits. This method is tested on different types of architectures, which are shown in Table~\ref{table3}, containing 1-D, grid, and 2-D, as well as architectures with or without Hamiltonian paths.  Detailed parameters of these architectures are given in Appendices~\ref{Coupling} and~\ref{Parameter}. The demonstration results, as depicted in Fig.~\ref{figbv}, demonstrate that the LCNNS approach is applicable to a variety of architectures under NISQ, and the average fidelity optimization rate is 14.85\%. Furthermore, its applicability extends to circuits including single-qubit gates, such as Deutsch-Jozsa circuits, QFT circuits.
   
   \begin{table}[]
   	\centering
   	\renewcommand{\arraystretch}{1.5} 
   	\setlength{\tabcolsep}{1.1pt} 
   	\caption{Parameters for different types of Architectures.}
   	\begin{ruledtabular}
       \begin{tabular}{ccccc}
       	Architecture & Makers & Qubit & Layout &Hamiltonian paths \\ \cline{0-4}
        Wuyuan II \footnotemark[1]   &Origin   & 6     & 1-D    & YES   \\
        Wukong \footnotemark[1]      &Origin   & 12    & 2-D    & NO   \\
        ScQ-10 \footnotemark[1]      &Quafu    & 10     & 1-D    & YES   \\
        ScQ-18  \footnotemark[1]     &Quafu    & 18    & 1-D    & YES   \\
       	Manila  \footnotemark[2]      &IBMQ     & 5     & 1-D    & YES                       \\
       	Quito  \footnotemark[2]       &IBMQ     & 5     & 2-D    & NO                        \\
       	Jakarta \footnotemark[2]      &IBMQ     & 7     & 2-D    & NO                        \\
       	Guadalupe \footnotemark[2]    &IBMQ     & 16    & 2-D    & NO                        \\
       	Tokyo  \footnotemark[2]       &IBMQ     & 20    & Grid   & YES                        \\
       	Almaden  \footnotemark[2]     &IBMQ     & 20    & Grid   & NO                        \\
       	Kolkata  \footnotemark[2]     &IBMQ     & 27    & 2-D    & NO                        \\ 
       \end{tabular}
    \end{ruledtabular}
     \footnotetext[1]{These QPUs are provided by the OriginQ Quantum Cloud Platform and the Quafu Cloud Platform in China, respectively.}
     \footnotetext[2]{These QPUs are provided by the IBMQ Cloud Platform.}
   	\label{table3}
   \end{table}

   	\pgfplotstableread[row sep=\\,col sep=&]{
   	interval & carT & carD  \\
   	Wuyuan II        & 0.908  & 0.899 \\
   	Wukong          & 0.82   & 0.833 \\
   	Manila          & 0.9167 & 0.9328   \\
   	Quito           & 0.6448 & 0.7033    \\
   	Jakarta         & 0.7487 & 0.9160 \\
   	Guadalupe       & 0.4972 & 0.5235  \\
   	Tokyo           & 0.1623 & 0.2549  \\
   	Almaden         & 0.2609 & 0.2870  \\
   	Kolkata         & 0.2420 &  0.3502  \\
   }\mydataobv
   
   \begin{figure}[]
   	\begin{adjustbox}{width=0.48\textwidth}
   		\begin{tikzpicture}
   			\begin{axis}[
   				ybar,
   				bar width=.5cm,
   				width=16cm,
   				height=.5\textwidth,
   				legend style={at={(0.99,0.98)},anchor=north east,cells={anchor=west}},
   				symbolic x coords={Wuyuan II,Wukong, Manila, Quito,Jakarta,Guadalupe,Tokyo,Almaden,Kolkata},
   				xtick=data,
   				nodes near coords,
   				nodes near coords align={vertical},
   				ymin=0,ymax=1,
   				title = {},
   				ylabel={Fidelity},
   				]
   				\addplot table[x=interval,y=carT]{\mydataobv};
   				\addplot table[x=interval,y=carD]{\mydataobv};
   				\legend{Compiler,LCNNS}
   			\end{axis}
   		\end{tikzpicture}
   	\end{adjustbox}
   	\caption{Comparison of the fidelity of the Bernstein-Vazirani algorithm for each architecture in Table~\ref{table3}. The fidelity calculation method is ESP. The  circuits demonstrated on the Origin and Quafu platforms are executed on the actual devices. Since the fidelity of the compiler execution on the Quafu platform is 0.0195 and 0.0008, respectively, and 0.326 and 0.011 for LCNNS, respectively, the data are too small to be presented in Fig~\ref{figbv}. And the circuits demonstrated on the IBMQ platform are executed on the fake\_provider simulator. This observation demonstrates the versatility of the LCNNS method, as it applies effectively to various architectural configurations, including 1-D, 2-D, and grid architectures, both with and without Hamiltonian paths. Notably, the LCNNS method exhibits superior fidelity performance compared with alternative methods.}
   	\label{figbv}
   \end{figure}
   
   \subsection{Data analysis}

   	\begin{figure*}[]
   	\subfigure[]{
   		\begin{adjustbox}{width=0.312\textwidth}
   			\includegraphics{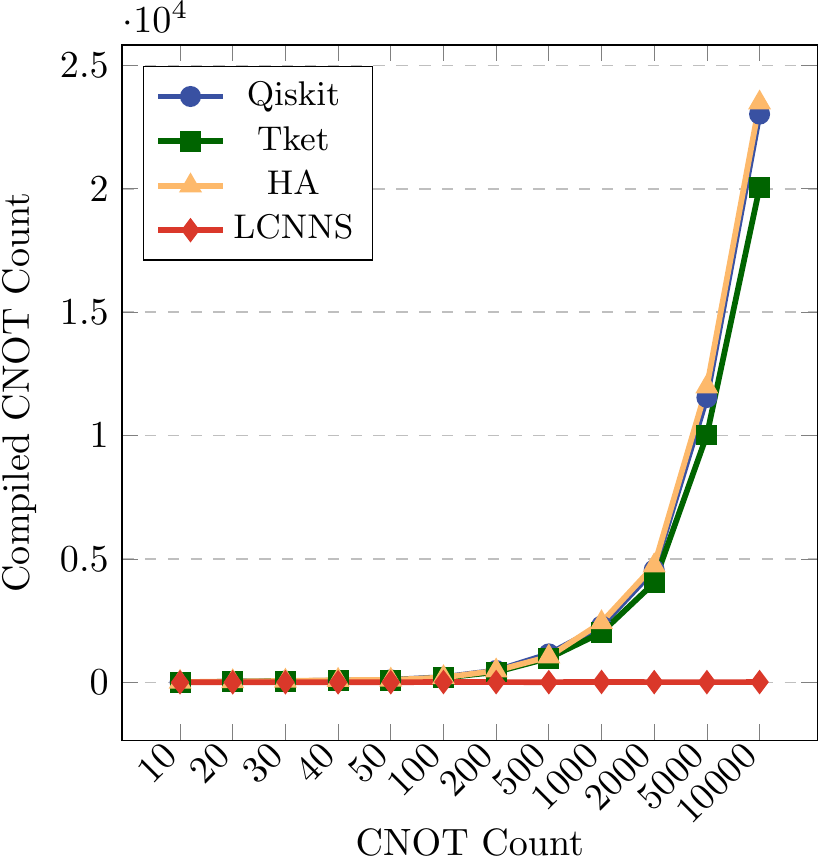}
   		\end{adjustbox}
   	}
   	\subfigure[]{
   		\begin{adjustbox}{width=0.312\textwidth}
   		\includegraphics{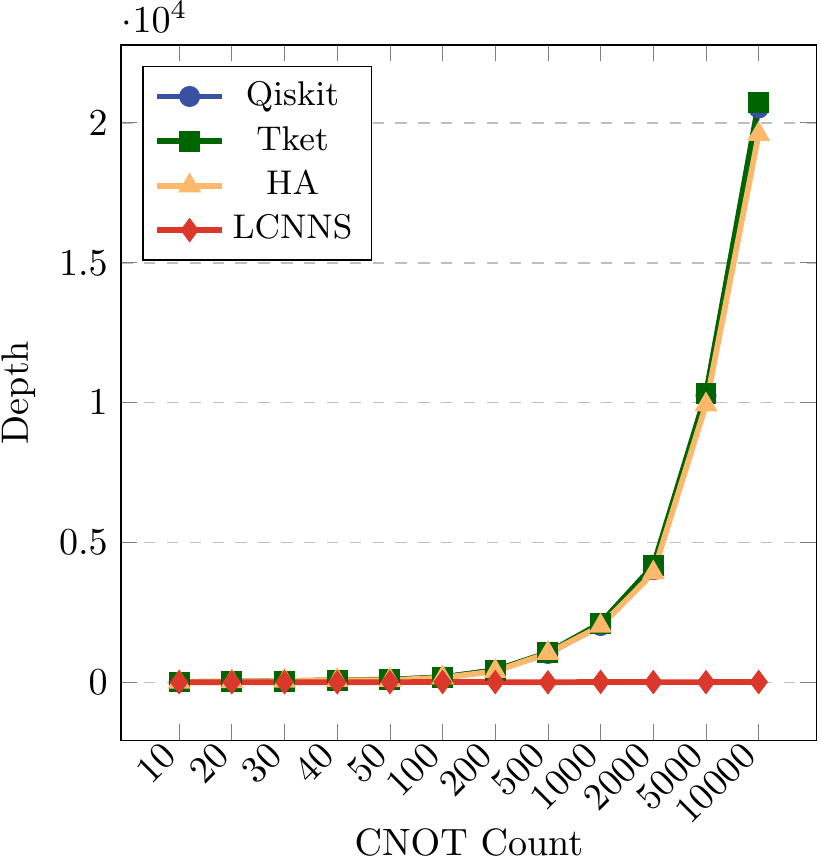}
   		\end{adjustbox}
   	}
   	\subfigure[]{
   		\begin{adjustbox}{width=0.312\textwidth}
   		\includegraphics{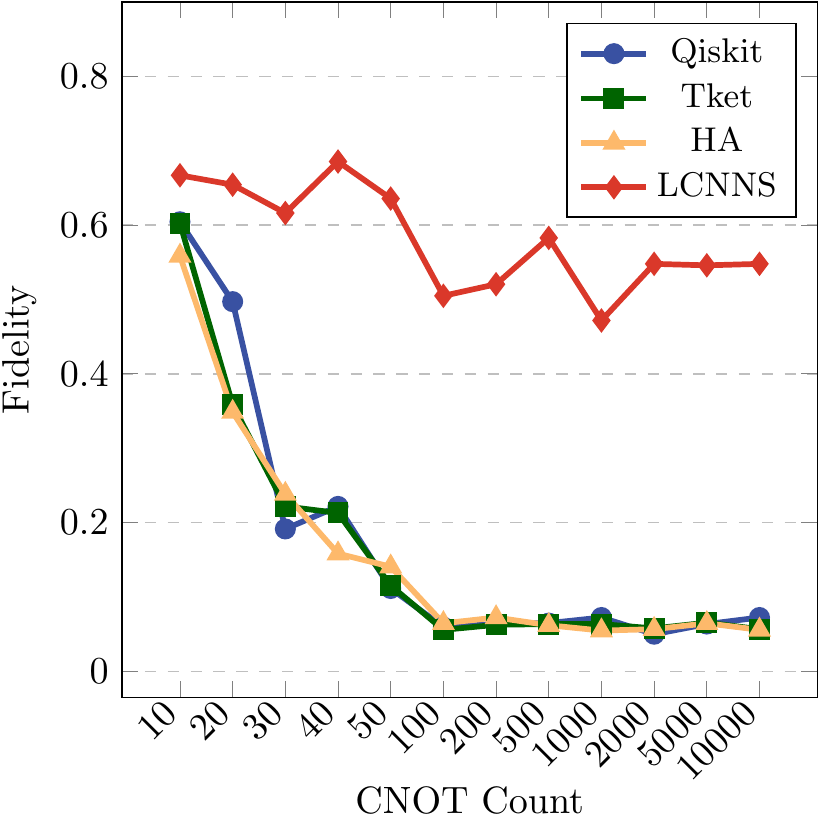}
   		\end{adjustbox}
   	}
   	\caption{Comparison of results in Table~\ref{table1} under three performance indicators. (a) Comparison of CNOT gate numbers: As the number of CNOT gates increases, the number of CNOT gates in the circuits after the NN of the comparison methods tends to increase. The LCNNS method stands out by maintaining a significantly lower count of CNOT gates compared to other methods.  (b) Comparison of depth: Similar to (a), the aspect of circuit depth follows the same trend, with LCNNS consistently achieving shallower circuit depth than other methods. (c) Comparison of fidelity: The increase in the number of CNOT gates results in a noticeable decrease in fidelity in circuits after applying the NN methods of comparison. However, the LCNNS method demonstrates the capability to maintain high fidelity even with an increased number of CNOT gates.}
   	\label{fig12}
   \end{figure*}

   \begin{figure*}[]
   	\subfigure[]{
   		\begin{adjustbox}{width=0.312\textwidth}
   			\includegraphics{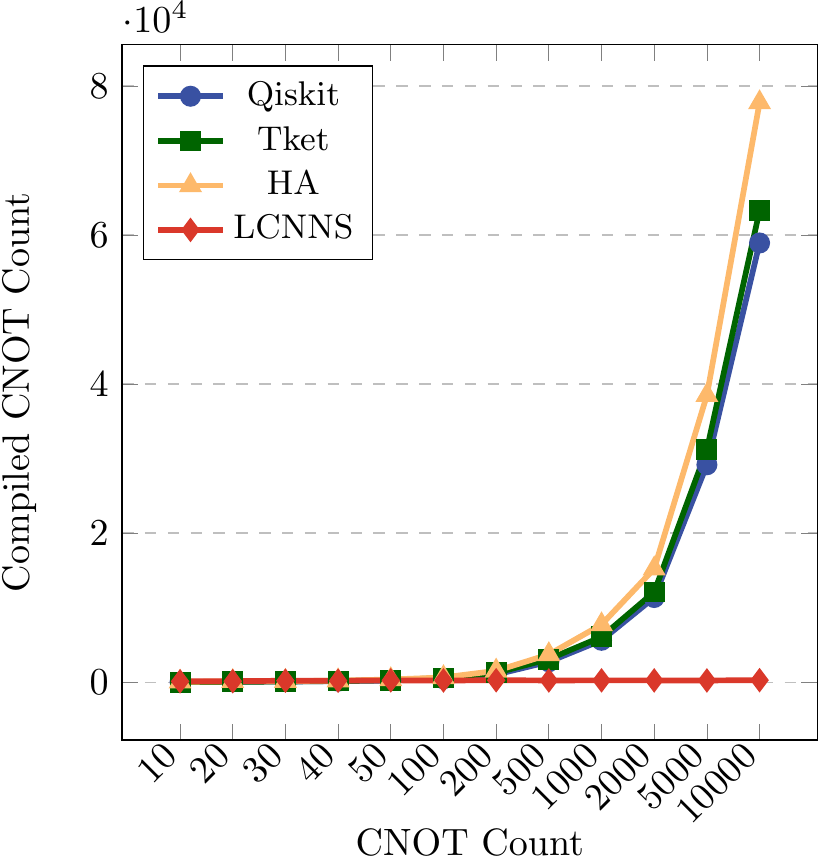}
   		\end{adjustbox}
   	}
   	\subfigure[]{
   		\begin{adjustbox}{width=0.312\textwidth}
   			\includegraphics{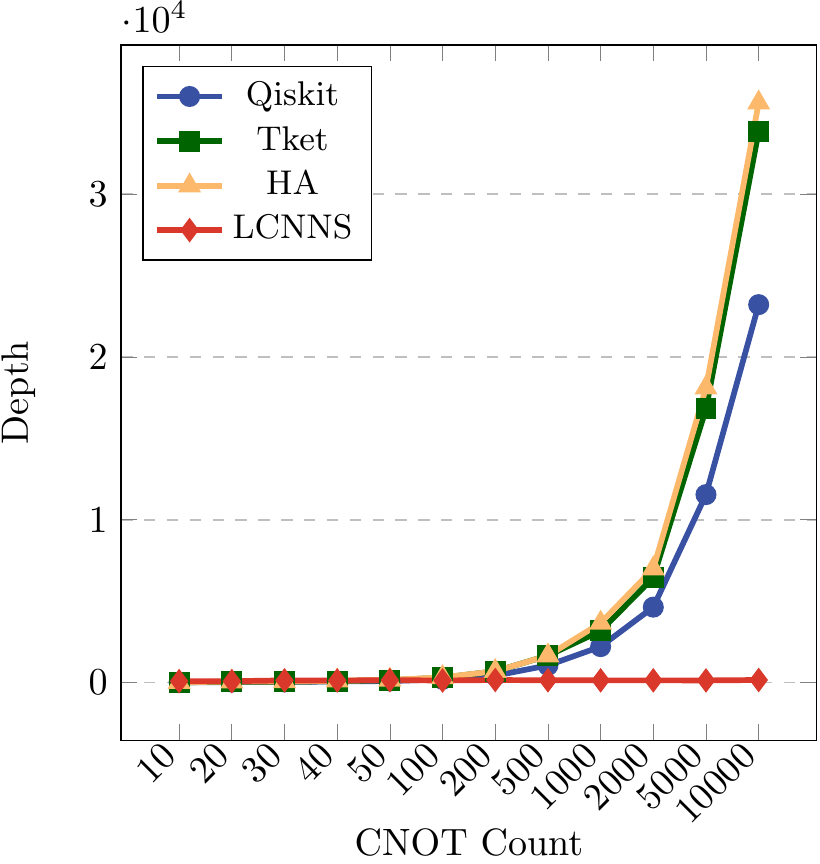}
   		\end{adjustbox}
   	}
   	\subfigure[]{
   		\begin{adjustbox}{width=0.312\textwidth}
   		\includegraphics{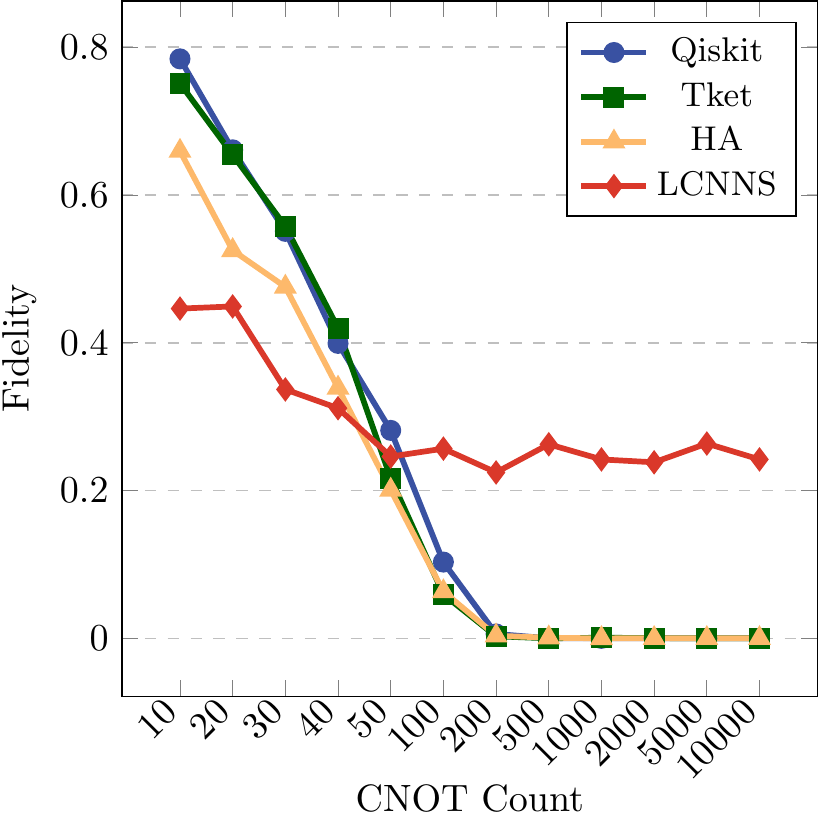}
   		\end{adjustbox}
   	}
   	\caption{Comparison of circuits in Table~\ref{table2} under three performance indicators. (a) Comparison of CNOT gate numbers: When the circuit size is large, the LCNNS method still keeps the number of CNOT gates at a low level, while the number of CNOT gates after NN of the comparison method grows linearly with the size. (b) Comparison of depth: The increase in the number of CNOT gates similarly leads to an increase in depth, which maintains essentially the same rate of increase in (a). (c) Comparison of fidelity: The fidelity of the comparison methods tends towards 0, but the fidelity of LCNNS overwhelmingly still manages to be greater than the 0.5 threshold.}
   \label{fig13}
   \end{figure*}

   \subsubsection{CNOT count analysis}
   The number of CNOT gates is a critical metric for assessing the complexity of a CNOT quantum circuit, and it has a significant impact on the efficiency and reliability of quantum computing. A higher number of CNOT gates indicates more complex interactions between qubits, leading to longer execution times and higher error rates. Therefore, reducing the number of CNOT gates after circuit transformation is essential for achieving efficient quantum computing. In contrast to the traditional way of inserting SWAP gates to achieve the NN of quantum gates, this paper achieves the NN synthesis of CNOT circuits by means of matrix transformation. This method enables the NN constraint of each gate in a CNOT circuit without inserting additional SWAP gates. As a result, it effectively reduces the number of CNOT gates in the circuit after the NN synthesis. Compared with Qiskit and HA, Tket further decomposes the CNOT gate and other gates into single-qubit gates after the neighbors. So the number of CNOT gates in the Tket method is smaller than Qiskit and HA. As shown in Figs.~\ref{fig12}(a) and \ref{fig13}(a), compared with the four NN methods and initial CNOT gates, the number of CNOT gates of the proposed LCNNS method is lower, and the average optimization rate of the number of CNOT gates reaches 39.1\%. Furthermore, the LCNNS method takes architecture connectivity into account during the initial mapping process, which results in fewer synthesized CNOT gates in most circuits compared with the synthesis methods of the comparison models, as shown in Fig. \ref{figex1}. This advantage becomes even more pronounced in circuits with a higher gate count.
   
   \subsubsection{Depth analysis}
   Similar to the number of CNOT gates, the depth of a CNOT quantum circuit is also an important indicator of the complexity and reliability of the CNOT quantum circuit. The depth of a CNOT quantum circuit refers to the number of layers of CNOT gates in the CNOT quantum circuit. Figs.~\ref{fig12}(b) and~\ref{fig13}(b) show the comparison between the three methods and the LCNNS method in terms of circuit depth. From Figs.~\ref{fig12}(b) and \ref{fig13}(b), it can be seen that the circuit depth of the LCNNS method is much lower than those of the Qiskit, Tket and HA methods, with an average optimization rate of 20.5\%. This is due to the significant effect of reducing the number of CNOT gates in the LCNNS method compared with other compiler methods.
   
   \subsubsection{Fidelity analysis}
   The fidelity of a quantum circuit refers to the degree of similarity between the output result of the circuit and the desired result, which can reflect the accuracy and reliability of the circuit. In quantum computing, it is essential that the output result of a quantum circuit is as close as possible to the desired result, so fidelity is one of the most important indicators of the performance of a quantum circuit. Figs.~\ref{fig12}(c) and \ref{fig13}(c) present a comparison between the Qiskit, Tket, HA and our LCNNS method in terms of circuit fidelity. It can be seen that the proposed LCNNS method greatly outperforms other methods in optimizing the circuit execution fidelity. Similarly, once the number of CNOT gates reaches a certain size, the fidelity of LCNNS synthesized circuits exceeds that of other synthesis methods, as shown in Fig. \ref{figex1}. There are two reasons for this. On the one hand, the CNOT circuit synthesis method achieves quantum gates that satisfy the NN constraint with as few CNOT gates as possible, and reducing the number of CNOT gates similarly reduces the source of double quantum gate errors, thus improving the fidelity of the synthesized CNOT circuit. On the other hand, the fidelity of CNOT gates is taken into account for the initial mapping, and the CNOT gates are prioritized to act on the path with a lower error rate during the matrix transformation. The trade-off between the number of CNOT gates and the fidelity of the simulation is chosen to achieve the CNOT circuit synthesis at a lower cost. This also explains why, in certain circuits as shown in Fig. \ref{figex1}b, the LCNNS method achieves the highest fidelity compared with other synthesis methods, even though it has more CNOT gates than the others. Therefore, in terms of fidelity, the LCNNS method is more suitable for circuits with a high number of CNOT gates.


   \section{Discussion}
   \label{sec:discussion}
   The method presented in this paper exhibits applicability across various IBM quantum computer architectures, especially those without Hamiltonian paths. Consequently, it can be effectively employed on a diverse range of quantum computing platforms in the NISQ. It is worth emphasizing that the suggested approach extends its utility to quantum circuits that incorporate not only pure CNOT gates but also additional single-qubit gates, such as Clifford+T circuits. This versatility arises from the capability of reassembling a quantum circuit containing CNOT subcircuits into a form that satisfies the NN constraint. This is achieved by synthesizing the CNOT subcircuits and subsequently reassembling the circuit with single-qubit gates. 
   
   The performance of the proposed method for compiling Clifford+T gate set circuits under varying CNOT gate ratios is evaluated. Randomly generated 5-qubit quantum circuits are used, with gate counts (including both single-qubit and CNOT gates) ranging from 20 to 10,000, and the CNOT gate ratio in the circuits is set to range from 10\% to 100\%. In comparison to the Qiskit method, the number of CNOT gates is selected as the performance metric, and demonstrations are conducted on the IBMQ\_Quito simulator, with the results presented in Fig.~\ref{fig: cnot num}.
   
   \begin{figure*}
   	\centering
   	\subfigure[]{ \begin{adjustbox}{width=0.3\textwidth}
   			\begin{tikzpicture}
   				\begin{axis}[
   					width=10cm,
   					height=10cm,
   					xlabel={CNOT ratio},
   					ylabel={Compiled CNOT Count},
   					xtick=data,
   					xticklabels={10\%, 20\%, 30\%, 40\%, 50\%, 60\%, 70\%, 80\%, 90\%, 100\%},
   					xticklabel style={rotate=45, anchor=east},
   					legend pos=north west,
   					grid=major,
   					]
   					\addplot[
   					color=mycolor1,
   					mark=*,line width=1.5pt, mark size=2pt
   					]
   					coordinates {
   						(1,2) (2,4) (3,2) (4,11) (5,14) (6,18) (7,21) (8,32) (9,31) (10,34)
   					};
   					\addplot[
   					color=mycolor4,line width=1.5pt, mark size=2pt,
   					mark=diamond*,
   					]
   					coordinates {
   						(1,10) (2,20) (3,8) (4,48) (5,34) (6,23) (7,39) (8,18) (9,22) (10,17)
   					};
   					\legend{Qiskit, LCNNS}
   				\end{axis}
   			\end{tikzpicture}
   	\end{adjustbox}}
   	\subfigure[]{ \begin{adjustbox}{width=0.3\textwidth}
   			\begin{tikzpicture}
   				\begin{axis}[
   					width=10cm,
   					height=10cm,
   					xlabel={CNOT ratio},
   					ylabel={Compiled CNOT Count},
   					xtick=data,
   					xticklabels={10\%, 20\%, 30\%, 40\%, 50\%, 60\%, 70\%, 80\%, 90\%, 100\%},
   					xticklabel style={rotate=45, anchor=east},
   					legend pos=north west,
   					grid=major,
   					]
   					\addplot[
   					color=mycolor1,
   					mark=*,line width=1.5pt, mark size=2pt
   					]
   					coordinates {(1,3) (2,22) (3,17) (4,43) (5,42) (6,61) (7,62) (8,87) (9,95) (10,111)};
   					\addplot[
   					color=mycolor4,line width=1.5pt, mark size=2pt,
   					mark=diamond*,
   					]
   					coordinates {(1,10) (2,36) (3,58) (4,42) (5,71) (6,119) (7,56) (8,88) (9,87) (10,20)};
   					\legend{Qiskit, LCNNS}
   				\end{axis}
   			\end{tikzpicture}
   	\end{adjustbox}}
   	\subfigure[]{ \begin{adjustbox}{width=0.3\textwidth}
   			\begin{tikzpicture}
   				\begin{axis}[
   					width=10cm,
   					height=10cm,
   					xlabel={CNOT ratio},
   					ylabel={Compiled CNOT Count},
   					xtick=data,
   					xticklabels={10\%, 20\%, 30\%, 40\%, 50\%, 60\%, 70\%, 80\%, 90\%, 100\%},
   					xticklabel style={rotate=45, anchor=east},
   					legend pos=north west,
   					grid=major,
   					]
   					\addplot[
   					color=mycolor1,
   					mark=*,line width=1.5pt, mark size=2pt
   					]
   					coordinates {(1,16) (2,39) (3,58) (4,94) (5,103) (6,145) (7,149) (8,167) (9,213) (10,238)};
   					\addplot[
   					color=mycolor4,line width=1.5pt, mark size=2pt,
   					mark=diamond*,
   					]
   					coordinates {(1,38) (2,67) (3,87) (4,119) (5,166) (6,187) (7,167) (8,176) (9,61) (10,21)};
   					\legend{Qiskit, LCNNS}
   				\end{axis}
   			\end{tikzpicture}
   	\end{adjustbox}}
   	\subfigure[]{ \begin{adjustbox}{width=0.3\textwidth}
   			\begin{tikzpicture}
   				\begin{axis}[
   					width=10cm,
   					height=10cm,
   					xlabel={CNOT ratio},
   					ylabel={Compiled CNOT Count},
   					xtick=data,
   					xticklabels={10\%, 20\%, 30\%, 40\%, 50\%, 60\%, 70\%, 80\%, 90\%, 100\%},
   					xticklabel style={rotate=45, anchor=east},
   					legend pos=north west,
   					grid=major,
   					]
   					\addplot[
   					color=mycolor1,
   					mark=*,line width=1.5pt, mark size=2pt
   					]
   					coordinates {(1,31) (2,57) (3,135) (4,172) (5,214) (6,279) (7,355) (8,390) (9,420) (10,443)};
   					\addplot[
   					color=mycolor4,line width=1.5pt, mark size=2pt,
   					mark=diamond*,
   					]
   					coordinates {(1,59) (2,120) (3,219) (4,224) (5,319) (6,356) (7,336) (8,334) (9,236) (10,18)};
   					\legend{Qiskit, LCNNS}
   				\end{axis}
   			\end{tikzpicture}
   	\end{adjustbox}}
   	\subfigure[]{ \begin{adjustbox}{width=0.3\textwidth}
   			\begin{tikzpicture}
   				\begin{axis}[
   					width=10cm,
   					height=10cm,
   					xlabel={CNOT ratio},
   				ylabel={Compiled CNOT Count},
   					xtick=data,
   					xticklabels={10\%, 20\%, 30\%, 40\%, 50\%, 60\%, 70\%, 80\%, 90\%, 100\%},
   					xticklabel style={rotate=45, anchor=east},
   					legend pos=north west,
   					grid=major,
   					]
   					\addplot[
   					color=mycolor1,
   					mark=*,line width=1.5pt, mark size=2pt
   					]
   					coordinates {(1,106) (2,253) (3,385) (4,442) (5,561) (6,690) (7,869) (8,902) (9,1011) (10,1155)};
   					\addplot[
   					color=mycolor4,line width=1.5pt, mark size=2pt,
   					mark=diamond*,
   					]
   					coordinates {(1,187) (2,331) (3,528) (4,725) (5,710) (6,885) (7,762) (8,843) (9,599) (10,21)};
   					\legend{Qiskit, LCNNS}
   				\end{axis}
   			\end{tikzpicture}
   	\end{adjustbox}}
   	\subfigure[]{ \begin{adjustbox}{width=0.3\textwidth}
   			\begin{tikzpicture}
   				\begin{axis}[
   					width=10cm,
   					height=10cm,
   					xlabel={CNOT ratio},
   				ylabel={Compiled CNOT Count},
   					xtick=data,
   					xticklabels={10\%, 20\%, 30\%, 40\%, 50\%, 60\%, 70\%, 80\%, 90\%, 100\%},
   					xticklabel style={rotate=45, anchor=east},
   					legend pos=north west,
   					grid=major,
   					]
   					\addplot[
   					color=mycolor1,
   					mark=*,line width=1.5pt, mark size=2pt
   					]
   					coordinates {(1,229) (2,467) (3,705) (4,908) (5,1116) (6,1362) (7,1622) (8,1889) (9,2098) (10,2275)};
   					\addplot[
   					color=mycolor4,line width=1.5pt, mark size=2pt,
   					mark=diamond*,
   					]
   					coordinates {(1,352) (2,667) (3,1121) (4,1277) (5,1577) (6,1668) (7,1817) (8,1662) (9,1082) (10,16)};
   					\legend{Qiskit, LCNNS}
   				\end{axis}
   			\end{tikzpicture}
   	\end{adjustbox}}
   	\subfigure[]{ \begin{adjustbox}{width=0.3\textwidth}
   			\begin{tikzpicture}
   				\begin{axis}[
   					width=10cm,
   					height=10cm,
   					xlabel={CNOT ratio},
   				ylabel={Compiled CNOT Count},
   					xtick=data,
   					xticklabels={10\%, 20\%, 30\%, 40\%, 50\%, 60\%, 70\%, 80\%, 90\%, 100\%},
   					xticklabel style={rotate=45, anchor=east},
   					legend pos=north west,
   					grid=major,
   					]
   					\addplot[
   					color=mycolor1,
   					mark=*,line width=1.5pt, mark size=2pt
   					]
   					coordinates {(1,486) (2,986) (3,1456) (4,1883) (5,2382) (6,2830) (7,3340) (8,3618) (9,4163) (10,4536)};
   					\addplot[
   					color=mycolor4,line width=1.5pt, mark size=2pt,
   					mark=diamond*,
   					]
   					coordinates {(1,796) (2,1464) (3,1966) (4,2477) (5,2919) (6,3382) (7,3423) (8,3327) (9,2457) (10,16)};
   					\legend{Qiskit, LCNNS}
   				\end{axis}
   			\end{tikzpicture}
   	\end{adjustbox}}
   	\subfigure[]{ \begin{adjustbox}{width=0.3\textwidth}
   			\begin{tikzpicture}
   				\begin{axis}[
   					width=10cm,
   					height=10cm,
   					xlabel={CNOT ratio},
   				ylabel={Compiled CNOT Count},
   					xtick=data,
   					xticklabels={10\%, 20\%, 30\%, 40\%, 50\%, 60\%, 70\%, 80\%, 90\%, 100\%},
   					xticklabel style={rotate=45, anchor=east},
   					legend pos=north west,
   					grid=major,
   					]
   					\addplot[
   					color=mycolor1,
   					mark=*,line width=1.5pt, mark size=2pt
   					]
   					coordinates {(1,1150) (2,2505) (3,3745) (4,4902) (5,5913) (6,7147) (7,8291) (8,9494) (9,10377) (10,11335)};
   					\addplot[
   					color=mycolor4,line width=1.5pt, mark size=2pt,
   					mark=diamond*,
   					]
   					coordinates {(1,1809) (2,3515) (3,5393) (4,6623) (5,7513) (6,8526) (7,8705) (8,8340) (9,5826) (10,18)};
   					\legend{Qiskit, LCNNS}
   				\end{axis}
   			\end{tikzpicture}
   	\end{adjustbox}}
   	\subfigure[]{ \begin{adjustbox}{width=0.3\textwidth}
   			\begin{tikzpicture}
   				\begin{axis}[
   					width=10cm,
   					height=10cm,
   					xlabel={CNOT ratio},
   				ylabel={Compiled CNOT Count},
   					xtick=data,
   					xticklabels={10\%, 20\%, 30\%, 40\%, 50\%, 60\%, 70\%, 80\%, 90\%, 100\%},
   					xticklabel style={rotate=45, anchor=east},
   					legend pos=north west,
   					grid=major,
   					]
   					\addplot[
   					color=mycolor1,
   					mark=*,line width=1.5pt, mark size=2pt
   					]
   					coordinates {(1,2501) (2,4938) (3,7464) (4,9664) (5,12218) (6,14204) (7,17066) (8,18835) (9,21170) (10,22943)};
   					\addplot[
   					color=mycolor4,line width=1.5pt, mark size=2pt,
   					mark=diamond*,
   					]
   					coordinates {(1,3734) (2,7240) (3,10440) (4,12873) (5,15774) (6,17079) (7,17744) (8,15927) (9,11708) (10,20)};
   					\legend{Qiskit, LCNNS}
   				\end{axis}
   			\end{tikzpicture}
   	\end{adjustbox}}
   	
   	\caption{Comparison of the number of CNOT gates between Qiskit and LCNNS methods under different CNOT gate ratios for various total circuit gate scales: (a)-(i) corresponding to 20, 50, 100, 200, 500, 1000, 2000, 5000, 10000 gates.}
   	\label{fig: cnot num}
   \end{figure*}

   From the results shown in Fig.~\ref{fig: cnot num}, it can be observed that with the increase in the CNOT gate ratio, the number of CNOT gates in the Qiskit method exhibits an upward trend, whereas the number of CNOT gates in the LCNNS method initially increases and then decreases. Intersection points in the CNOT gate counts between the two methods occur at certain ratios. This indicates that the LCNNS method is more suitable for handling circuits with a higher number of CNOT gates. As the CNOT gate ratio increases, the advantage of the LCNNS method becomes more apparent, with the number of CNOT gates showing an initial increase followed by a decrease. Therefore, when the CNOT gate ratio reaches a certain threshold, the optimization effect of the LCNNS method surpasses that of the Qiskit method.

   To further investigate the CNOT gate ratio threshold at which the LCNNS method outperforms the Qiskit method for different numbers of qubits, tests are conducted on 5-qubit, 7-qubit, 14-qubit, 16-qubit, and 20-qubit circuits, each with 1,000 gates. The demonstrations of these circuits were conducted on IBMQ\_Quito, BMQ\_Jakarta, BMQ\_Melbourne, BMQ\_Guadalupe, and BMQ\_Almaden simulators, respectively. The results, shown in Fig.~\ref{fig:qubit_threshold}, indicate that as the number of qubits increases, the threshold at which the LCNNS method outperforms the Qiskit method gradually approaches 100\%. It is important to note that Fig.~\ref{fig:qubit_threshold} only shows the threshold for circuits with 1,000 gates, and the thresholds may differ for circuits with different gate counts. This phenomenon provides substantial insights for quantum circuit compilation: when the number of CNOT gates in a circuit is low or below a certain threshold, the insertion of SWAP gates to achieve nearest-neighbor compliance may be employed; conversely, when the number of CNOT gates is high or exceeds a certain threshold, utilizing the nearest-neighbor synthesis for CNOT gates might be more beneficial.
   
   \begin{figure}
   	\centering
   	\begin{adjustbox}{width=0.4\textwidth}
   		\begin{tikzpicture}
   			\begin{axis}[
   				xlabel={Number of Qubits},
   				ylabel={Threshold (\%)},
   				xmin=0, xmax=25,
   				ymin=70, ymax=100,
   				xtick={0, 5, 10, 15, 20, 25},
   				ytick={70, 75, 80, 85, 90, 95, 100},
   				legend pos=south east,
   				grid=major,
   				width=\textwidth,
   				height=0.618\textwidth,
   				xlabel style={font=\Large},
   				ylabel style={font=\Large},
   				tick label style={font=\Large},
   				legend style={font=\Large}
   				]
   				\addplot[
   				color=mycolor1,mark=pentagon*,line width=2pt, mark size=5pt, dashed 	]
   				coordinates {
   					(5,74) (7,82) (14,96) (16,97) (20,99)
   				};
   				\legend{Threshold}
   			\end{axis}
   		\end{tikzpicture}
   	\end{adjustbox}
   	\caption{Relationship between the number of qubits and threshold. The threshold means that when the ratio of CNOT gates in the circuit is greater than this value, the LCNNS method will outperform Qiskit.}
   	\label{fig:qubit_threshold}
   \end{figure}

In summary, this section reveals the significant advantages of the LCNNS method in handling quantum circuits with high CNOT gate ratios and provides new optimization strategies for quantum circuit compilation. As the CNOT gate ratio increases, the LCNNS method demonstrates stronger optimization capabilities, making it a preferable choice over the traditional Qiskit method under certain conditions. The implication is that CNOT circuits generated using this method can be easily scaled to more complex quantum circuits, substantially enhancing the fidelity of CNOT circuits post-synthesis and execution. This enhancement is of paramount significance for the practical deployment of CNOT circuits in various quantum computing contexts. Future work could explore optimization strategies for quantum circuits of different scales and structures, further enhancing quantum circuit compilation techniques.
   
   \section{Conclusion}
   \label{sec:conclusion}
   This work presents an effective approach to synthesizing CNOT quantum circuits on general quantum computing devices with and without Hamiltonian paths. Specifically, the proposed key-qubit priority model and tabu search algorithm are employed to dynamically adjust the mapping method of key qubits, with the aim of reducing the number of CNOT gates and increasing circuit fidelity after matrix transformation. Additionally, a noise-aware NN synthesis method based on layer convergence is presented. These approaches effectively improve the fidelity of the synthesized circuit, providing a new solution to the problem of synthesizing CNOT circuits on real quantum computing devices. The demonstration results, performed on the IBMQ\_Quito quantum architecture and other architectures in the cloud platform, demonstrate that the proposed method can significantly enhance the execution fidelity of CNOT quantum circuits on real quantum hardware. While these methods enhance fidelity significantly, it is crucial to note that the complexity of synthesizing circuits increases factorially with the number of qubits, posing scalability challenges. Future work will focus on developing more efficient synthesis techniques to address this complexity and extend our methods to more complex quantum circuits.

	\section*{Acknowledgment}
    The work was supported by the National Natural Science Foundation of China under Grant number 62072259, in part by the Natural Science Foundation of Jiangsu Province under Grant number BK20221411, in part by 
the Natural Science Foundation of Nantong under Grant number JC2024100, in part by the PhD Start-up Fund of Nantong University under Grant number 23B03, in part by the Postgraduate Research and Practice Innovation Program of Jiangsu Province under Grant number SJCX21\_1448 and SJCX23\_1782.

	\section*{Data and code availability}
	The raw test data of this study and the software codes for generating the test data are publicly available at https://github.com/M-qiangZhu/Nearest-neighbor-synthesis-of-CNOT-circuits-on-general-quantum-architectures.


\bibliography{ref}

\appendix

\section{Coupling graphs}
\label{Coupling}
In this Appendix, we show various architectures on different quantum cloud platforms. These platforms include the OriginQ Quantum Cloud Platform~\cite{origin} in China, the Cloud Platform of Beijing Institute of Quantum Information Science~\cite{quafu} in China, and the IBMQ Cloud Platform. 

OriginQ Quantum Cloud Platform is launched by OriginQ Quantum Computing Technology (Hefei) Co. It provides real quantum computing, quantum computing simulators, and super-quantum hybrid computing back-end access. It provides basic software and development tools for developers, and industry application services for advanced companies in biochemistry, financial technology, big data, and machine learning. Users can access the following link, https://qcloud.originqc.com.cn/zh/home. The coupling graph of the quantum computing device provided by OriginQ is shown in Fig.~\ref{originq}.

 \begin{figure}[h]
	\centering
	\subfigure[]{\begin{adjustbox}{width=0.4\textwidth}
			\begin{tikzpicture}[>=stealth]
				\node (Q0) at (-4,0) [circle,draw=black!100,inner sep=2.2mm] {\Large0};
				\node (Q1) at (-2,0) [circle,draw=black!100,inner sep=2.2mm] {\Large1};
				\node (Q2) at (0,0) [circle,draw=black!100,inner sep=2.2mm] {\Large2};
				\node (Q3) at (2,0) [circle,draw=black!100,inner sep=2.2mm] {\Large3};
				\node (Q4) at (4,0) [circle,draw=black!100,inner sep=2.2mm] {\Large4};
				\node (Q5) at (6,0) [circle,draw=black!100,inner sep=2.2mm] {\Large5};
				\draw [thick] (Q0) to (Q1);
				\draw [thick] (Q1) to (Q2);
				\draw [thick] (Q2) to (Q3);
				\draw [thick] (Q3) to (Q4);
				\draw [thick] (Q4) to (Q5);
			\end{tikzpicture}
	\end{adjustbox}}
	\subfigure[]{
	\begin{adjustbox}{width=0.4\textwidth}
		\begin{tikzpicture}
			\node (Q1) at (-4,10) [circle,dashed,draw=black!100,inner sep=2.2mm] {\Large1};
			\node (Q2) at (-2,10) [circle,dashed,draw=black!100,inner sep=2.2mm] {\Large2};
			\node (Q3) at (0,10) [circle,dashed,draw=black!100,inner sep=2.2mm] {\Large3};
			\node (Q4) at (2,10) [circle,dashed,draw=black!100,inner sep=2.2mm] {\Large4};
			\node (Q5) at (4,10) [circle,dashed,draw=black!100,inner sep=2.2mm] {\Large5};
			\node (Q6) at (6,10) [circle,dashed,draw=black!100,inner sep=2.2mm] {\Large6};
			\node (Q7) at (-4,8) [circle,dashed,draw=black!100,inner sep=2.2mm] {\Large7};
			\node (Q8) at (-2,8) [circle,dashed,draw=black!100,inner sep=2.2mm] {\Large8};
			\node (Q9) at (0,8)  [circle,dashed,draw=black!100,inner sep=2.2mm] {\Large9};
			\node (Q10) at (2,8) [circle,dashed,draw=black!100,inner sep=1.8mm] {\Large10};
			\node (Q11) at (4,8) [circle,dashed,draw=black!100,inner sep=1.8mm] {\Large11};
			\node (Q12) at (6,8) [circle,dashed,draw=black!100,inner sep=1.8mm] {\Large12};
			\foreach \i in {13,...,18} {
				\pgfmathtruncatemacro{\label}{\i}
				\node (Q\i) at (2*\i-30, 6) [circle, dashed, draw=black!100, inner sep=1.8mm] {\Large\label};
			}
			\foreach \i in {19,...,24} {
				\pgfmathtruncatemacro{\label}{\i}
				\node (Q\i) at (2*\i-42, 4) [circle, dashed, draw=black!100, inner sep=1.8mm] {\Large\label};
			}
			\foreach \i in {25,...,30} {
				\pgfmathtruncatemacro{\label}{\i}
				\node (Q\i) at (2*\i-54, 2) [circle, dashed, draw=black!100, inner sep=1.8mm] {\Large\label};
			}
			\foreach \i in {31,...,36} {
				\pgfmathtruncatemacro{\label}{\i}
				\node (Q\i) at (2*\i-66, 0) [circle, dashed, draw=black!100, inner sep=1.8mm] {\Large\label};
			}
			\foreach \i in {37,...,39} {
				\pgfmathtruncatemacro{\label}{\i}
				\node (Q\i) at (2*\i-78, -2) [circle, dashed, draw=black!100, inner sep=1.8mm] {\Large\label};
			}
			\node (Q40) at (2,-2) [circle,draw=black!100,inner sep=1.8mm] {\Large40};
			\node (Q41) at (4,-2) [circle,dashed,draw=black!100,inner sep=1.8mm] {\Large41};
			\node (Q42) at (6,-2) [circle,dashed,draw=black!100,inner sep=1.8mm] {\Large42};
			\foreach \i in {43,...,44} {
				\pgfmathtruncatemacro{\label}{\i}
				\node (Q\i) at (2*\i-90, -4) [circle, dashed, draw=black!100, inner sep=1.8mm] {\Large\label};
			}
			\node (Q45) at (0,-4) [circle,draw=black!100,inner sep=1.8mm] {\Large45};
			\node (Q46) at (2,-4) [circle,draw=black!100,inner sep=1.8mm] {\Large46};
			\node (Q47) at (4,-4) [circle,dashed,draw=black!100,inner sep=1.8mm] {\Large47};
			\node (Q48) at (6,-4) [circle,draw=black!100,inner sep=1.8mm] {\Large48};
			\foreach \i in {49,...,51} {
				\pgfmathtruncatemacro{\label}{\i}
				\node (Q\i) at (2*\i-102, -6) [circle, dashed, draw=black!100, inner sep=1.8mm] {\Large\label};
			}
			\foreach \i in {52,...,54} {
				\pgfmathtruncatemacro{\label}{\i}
				\node (Q\i) at (2*\i-102, -6) [circle, draw=black!100, inner sep=1.8mm] {\Large\label};
			}
			\foreach \i in {55,...,58} {
				\pgfmathtruncatemacro{\label}{\i}
				\node (Q\i) at (2*\i-114, -8) [circle, dashed,draw=black!100, inner sep=1.8mm] {\Large\label};
			}
			\node (Q59) at (4,-8) [circle,draw=black!100,inner sep=1.8mm] {\Large59};
			\node (Q60) at (6,-8) [circle,draw=black!100,inner sep=1.8mm] {\Large60};
			\foreach \i in {61,...,64} {
				\pgfmathtruncatemacro{\label}{\i}
				\node (Q\i) at (2*\i-126, -10) [circle, dashed,draw=black!100, inner sep=1.8mm] {\Large\label};
			}
			\node (Q65) at (4,-10) [circle,draw=black!100,inner sep=1.8mm] {\Large65};
			\node (Q66) at (6,-10) [circle,draw=black!100,inner sep=1.8mm] {\Large66};
			\foreach \i in {67,...,70} {
				\pgfmathtruncatemacro{\label}{\i}
				\node (Q\i) at (2*\i-138, -12) [circle, dashed,draw=black!100, inner sep=1.8mm] {\Large\label};
			}
			\node (Q71) at (4,-12) [circle,draw=black!100,inner sep=1.8mm] {\Large71};
			\node (Q72) at (6,-12) [circle,dashed,draw=black!100,inner sep=1.8mm] {\Large72};
			\foreach \x in {1,2,3,4,5}{%
				\pgfmathtruncatemacro{\nextx}{\x + 1}
				\draw [dashed] (Q\x) to (Q\nextx);
			}
			\foreach \x in {7,...,11}{%
				\pgfmathtruncatemacro{\nextx}{\x + 1}
				\draw [dashed] (Q\x) to (Q\nextx);
			}
			\foreach \x in {13,...,17}{%
				\pgfmathtruncatemacro{\nextx}{\x + 1}
				\draw [dashed] (Q\x) to (Q\nextx);
			}
			\foreach \x in {19,...,23}{%
				\pgfmathtruncatemacro{\nextx}{\x + 1}
				\draw [dashed] (Q\x) to (Q\nextx);
			}
			\foreach \x in {25,...,29}{%
				\pgfmathtruncatemacro{\nextx}{\x + 1}
				\draw [dashed] (Q\x) to (Q\nextx);
			}
			\foreach \x in {31,...,35}{%
				\pgfmathtruncatemacro{\nextx}{\x + 1}
				\draw [dashed] (Q\x) to (Q\nextx);
			}
			\foreach \x in {37,...,41}{%
				\pgfmathtruncatemacro{\nextx}{\x + 1}
				\draw [dashed] (Q\x) to (Q\nextx);
			}
			\draw [dashed] (Q43) to (Q44);
			\draw [dashed] (Q44) to (Q45);
			\draw [thick] (Q45) to (Q46);
			\draw [dashed] (Q46) to (Q47);
			\draw [dashed] (Q47) to (Q48);
			\foreach \x in {49,...,51}{%
				\pgfmathtruncatemacro{\nextx}{\x + 1}
				\draw [dashed] (Q\x) to (Q\nextx);
			}
			\draw [thick] (Q52) to (Q53);
			\draw [thick] (Q53) to (Q54);
			\foreach \x in {55,...,58}{%
				\pgfmathtruncatemacro{\nextx}{\x + 1}
				\draw [dashed] (Q\x) to (Q\nextx);
			}
			\draw [thick] (Q59) to (Q60);
			\foreach \x in {61,...,64}{%
				\pgfmathtruncatemacro{\nextx}{\x + 1}
				\draw [dashed] (Q\x) to (Q\nextx);
			}
			\draw [thick] (Q65) to (Q66);
			\foreach \x in {67,...,71}{%
				\pgfmathtruncatemacro{\nextx}{\x + 1}
				\draw [dashed] (Q\x) to (Q\nextx);
			}
			\foreach \x in {1,7,13,19,25,31,37,43,49,55,61}{%
				\pgfmathtruncatemacro{\nextx}{\x + 6}
				\draw [dashed] (Q\x) to (Q\nextx);
			}
			\foreach \x in {2,8,14,...,62}{%
				\pgfmathtruncatemacro{\nextx}{\x + 6}
				\draw [dashed] (Q\x) to (Q\nextx);
			}       
			\foreach \x in {3,9,15,...,63}{%
				\pgfmathtruncatemacro{\nextx}{\x + 6}
				\draw [dashed] (Q\x) to (Q\nextx);
			}
			\foreach \x in {4,10,16,...,34}{%
				\pgfmathtruncatemacro{\nextx}{\x + 6}
				\draw [dashed] (Q\x) to (Q\nextx);
			}
			\foreach \x in {5,11,17,...,59}{%
				\pgfmathtruncatemacro{\nextx}{\x + 6}
				\draw [dashed] (Q\x) to (Q\nextx);
			}
			\draw [thick] (Q40) to (Q46);      
			\draw [thick] (Q46) to (Q52); 
			\foreach \x in {52,58,64}{%
				\pgfmathtruncatemacro{\nextx}{\x + 6}
				\draw [dashed] (Q\x) to (Q\nextx);
			}
			\draw [thick] (Q65) to (Q71); 
			\foreach \x in {6,12,18,...,42}{%
				\pgfmathtruncatemacro{\nextx}{\x + 6}
				\draw [dashed] (Q\x) to (Q\nextx);
			}
			\foreach \x in {48,54,60}{%
				\pgfmathtruncatemacro{\nextx}{\x + 6}
				\draw [thick] (Q\x) to (Q\nextx);
			}
			\draw [dashed] (Q66) to (Q72); 
		\end{tikzpicture}
	\end{adjustbox}}
	\caption{(a) The coupling graph of Wuyuan II. (b) The coupling graph of Wukong. The dashed qubits and connecting lines indicate unavailable qubits and couplings. Therefore, the number of qubits currently available for this QPU is 12.}
 \label{originq}
\end{figure}

Quafu is a quantum computing cloud platform launched by Beijing Institute of Quantum Information Science in collaboration with Institute of Physics, Chinese Academy of Sciences and Tsinghua University, with the meaning of "quantum future". At present, there are three superconducting quantum chips on the platform, which have 136, 18 and 10 qubits, respectively, and users can independently choose the appropriate chip to run quantum computing tasks. Users can access the following link, https://quafu.baqis.ac.cn. The coupling graph of the quantum computing device provided by Ouafu is shown in Fig.~\ref{quafu}.

 \begin{figure}[]
	\centering
	\subfigure[]{	\begin{adjustbox}{width=0.35\textwidth}
			\begin{tikzpicture}[>=stealth]
				\node (Q0) at (-4,0) [circle,draw=black!100,inner sep=2.2mm] {\Large0};
				\node (Q1) at (-2,0) [circle,draw=black!100,inner sep=2.2mm] {\Large1};
				\node (Q2) at (0,0) [circle,draw=black!100,inner sep=2.2mm] {\Large2};
				\node (Q3) at (2,0) [circle,draw=black!100,inner sep=2.2mm] {\Large3};
				\node (Q4) at (4,0) [circle,draw=black!100,inner sep=2.2mm] {\Large4};
				\node (Q5) at (4,-2) [circle,draw=black!100,inner sep=2.2mm] {\Large5};
				\node (Q6) at (2,-2) [circle,draw=black!100,inner sep=2.2mm] {\Large6};
				\node (Q7) at (0,-2) [circle,draw=black!100,inner sep=2.2mm] {\Large7};
				\node (Q8) at (-2,-2) [circle,draw=black!100,inner sep=2.2mm] {\Large8};
				\node (Q9) at (-4,-2) [circle,draw=black!100,inner sep=2.2mm] {\Large9};
				\draw [thick] (Q0) to (Q1);
				\draw [thick] (Q1) to (Q2);
				\draw [thick] (Q2) to (Q3);
				\draw [thick] (Q3) to (Q4);
				\draw [thick] (Q4) to (Q5);
				\draw [thick] (Q5) to (Q6);
				\draw [thick] (Q6) to (Q7);
				\draw [thick] (Q7) to (Q8);
				\draw [thick] (Q8) to (Q9);
			\end{tikzpicture}
\end{adjustbox}}
	\subfigure[]{
\begin{adjustbox}{width=0.35\textwidth}
	\begin{tikzpicture}[>=stealth]
		\node (Q0) at (0,0) [circle,draw=black!100,inner sep=2.2mm] {\Large0};
		\node (Q1) at (2,0) [circle,draw=black!100,inner sep=2.2mm] {\Large1};
		\node (Q2) at (4,0) [circle,draw=black!100,inner sep=2.2mm] {\Large2};
		\node (Q3) at (4,-2) [circle,draw=black!100,inner sep=2.2mm] {\Large3};
		\node (Q4) at (2,-2) [circle,draw=black!100,inner sep=2.2mm] {\Large4};
		\node (Q5) at (0,-2) [circle,draw=black!100,inner sep=2.2mm] {\Large5};
		\node (Q6) at (-2,-2) [circle,draw=black!100,inner sep=2.2mm] {\Large6};
		\node (Q7) at (-4,-2) [circle,draw=black!100,inner sep=2.2mm] {\Large7};
		\node (Q8) at (-4,-4) [circle,draw=black!100,inner sep=2.2mm] {\Large8};
		\node (Q9) at (-2,-4) [circle,draw=black!100,inner sep=2.2mm] {\Large9};
		\node (Q10) at (0,-4) [circle,draw=black!100,inner sep=1.5mm] {\Large10};
		\node (Q11) at (2,-4) [circle,draw=black!100,inner sep=1.5mm] {\Large11};
		\node (Q12) at (4,-4) [circle,draw=black!100,inner sep=1.5mm] {\Large12};
		\node (Q13) at (4,-6) [circle,draw=black!100,inner sep=1.5mm] {\Large13};
		\node (Q14) at (2,-6) [circle,draw=black!100,inner sep=1.5mm] {\Large14};
		\node (Q15) at (0,-6) [circle,draw=black!100,inner sep=1.5mm] {\Large15};
		\node (Q16) at (-2,-6) [circle,draw=black!100,inner sep=1.5mm] {\Large16};
		\node (Q17) at (-4,-6) [circle,draw=black!100,inner sep=1.5mm] {\Large17};
		\foreach \x in {0,1,2,3,4,5,6,7,8,9,10,11,12,13,14,15,16}{%
			\pgfmathtruncatemacro{\nextx}{\x + 1}
			\draw [thick] (Q\x) to (Q\nextx);
		}
	\end{tikzpicture}
\end{adjustbox}}
	
	\caption{(a) and (b) are the coupling graphs of the QPUs of the Quafu cloud platform at 10 and 18 qubits, respectively, and both are 1-D architectures.}
 \label{quafu}
\end{figure}

IBM Quantum Platform is a cloud-based quantum computing service provided by IBM that allows users to access and explore quantum computing over the Internet. It provides quantum computer hardware, the Qiskit programming framework, and a wealth of educational resources to provide researchers and developers with easy access to quantum computing resources and tools. IBMQ supports the development of quantum computing research and applications through open cloud services. Users can access the following link, https://quantum-computing.ibm.com. In addition to the architecture shown in Fig.~\ref{fig2}, the coupling graph of IBMQ's quantum computing devices used in the demonstration is shown in Fig.~\ref{ibmqu}.

 \begin{figure*}[]
	\centering
	\subfigure[]{
		\begin{adjustbox}{width=0.4\textwidth}
			\begin{tikzpicture}[>=stealth]
				\node (Q0) at (-4,0) [circle,draw=black!100,inner sep=2.2mm] {\Large0};
				\node (Q1) at (-2,0) [circle,draw=black!100,inner sep=2.2mm] {\Large1};
				\node (Q2) at (0,0) [circle,draw=black!100,inner sep=2.2mm] {\Large2};
				\node (Q3) at (2,0) [circle,draw=black!100,inner sep=2.2mm] {\Large3};
				\node (Q4) at (4,0) [circle,draw=black!100,inner sep=2.2mm] {\Large4};
				\draw [thick] (Q0) to (Q1);
				\draw [thick] (Q1) to (Q2);
				\draw [thick] (Q2) to (Q3);
				\draw [thick] (Q3) to (Q4);
			\end{tikzpicture}
		\end{adjustbox}
	}
    \subfigure[]{
    	\begin{adjustbox}{width=0.24\textwidth}
    		\begin{tikzpicture}[>=stealth]
    			\node (Q0) at (-4,0) [circle,draw=black!100,inner sep=2.2mm] {\Large0};
    			\node (Q1) at (-2,0) [circle,draw=black!100,inner sep=2.2mm] {\Large1};
    			\node (Q2) at (0,0) [circle,draw=black!100,inner sep=2.2mm] {\Large2};
    			\node (Q3) at (-2,-2) [circle,draw=black!100,inner sep=2.2mm] {\Large3};
    			\node (Q4) at (-4,-4) [circle,draw=black!100,inner sep=2.2mm] {\Large4};
    			\node (Q5) at (-2,-4) [circle,draw=black!100,inner sep=2.2mm] {\Large5};
    			\node (Q6) at (0,-4) [circle,draw=black!100,inner sep=2.2mm] {\Large6};
    			\draw [thick] (Q0) to (Q1);
    			\draw [thick] (Q1) to (Q2);
    			\draw [thick] (Q1) to (Q3);
    			\draw [thick] (Q3) to (Q5);
    			\draw [thick] (Q4) to (Q5);
    			\draw [thick] (Q5) to (Q6);
    		\end{tikzpicture}
    	\end{adjustbox}
    }
	\subfigure[]{	
			\begin{adjustbox}{width=0.4\textwidth}
			\begin{tikzpicture}[>=stealth]
				\node (Q0) at (0,6) [circle,draw=black!100,inner sep=2.2mm] {\Large0};
				\node (Q1) at (2,6) [circle,draw=black!100,inner sep=2.2mm] {\Large1};
				\node (Q2) at (4,6) [circle,draw=black!100,inner sep=2.2mm] {\Large2};
				\node (Q3) at (6,6) [circle,draw=black!100,inner sep=2.2mm] {\Large3};
				\node (Q4) at (8,6) [circle,draw=black!100,inner sep=2.2mm] {\Large4};
				\node (Q5) at (0,4) [circle,draw=black!100,inner sep=2.2mm] {\Large5};
				\node (Q6) at (2,4) [circle,draw=black!100,inner sep=2.2mm] {\Large6};
				\node (Q7) at (4,4) [circle,draw=black!100,inner sep=2.2mm] {\Large7};
				\node (Q8) at (6,4) [circle,draw=black!100,inner sep=2.2mm] {\Large8};
				\node (Q9) at (8,4) [circle,draw=black!100,inner sep=2.2mm] {\Large9};
				\node (Q10) at (0,2) [circle,draw=black!100,inner sep=1.8mm] {\Large10};
				\node (Q11) at (2,2) [circle,draw=black!100,inner sep=1.8mm] {\Large11};
				\node (Q12) at (4,2) [circle,draw=black!100,inner sep=1.8mm] {\Large12};
				\node (Q13) at (6,2) [circle,draw=black!100,inner sep=1.8mm] {\Large13};
				\node (Q14) at (8,2) [circle,draw=black!100,inner sep=1.8mm] {\Large14};
				\node (Q15) at (0,0) [circle,draw=black!100,inner sep=1.8mm] {\Large15};
				\node (Q16) at (2,0) [circle,draw=black!100,inner sep=1.8mm] {\Large16};
				\node (Q17) at (4,0) [circle,draw=black!100,inner sep=1.8mm] {\Large17};
				\node (Q18) at (6,0) [circle,draw=black!100,inner sep=1.8mm] {\Large18};
				\node (Q19) at (8,0) [circle,draw=black!100,inner sep=1.8mm] {\Large19};
				\draw [thick] (Q0) to (Q1);
				\draw [thick] (Q1) to (Q2);
				\draw [thick] (Q2) to (Q3);
				\draw [thick] (Q3) to (Q4);
				\draw [thick] (Q0) to (Q5);
				\draw [thick] (Q1) to (Q6);
				\draw [thick] (Q2) to (Q7);
				\draw [thick] (Q3) to (Q8);
				\draw [thick] (Q4) to (Q9);
				\draw [thick] (Q3) to (Q9);
				\draw [thick] (Q4) to (Q8);
				\draw [thick] (Q5) to (Q6);
				\draw [thick] (Q6) to (Q7);
				\draw [thick] (Q7) to (Q8);
				\draw [thick] (Q8) to (Q9);
				\draw [thick] (Q5) to (Q10);
				\draw [thick] (Q6) to (Q11);
				\draw [thick] (Q7) to (Q12);
				\draw [thick] (Q8) to (Q13);
				\draw [thick] (Q5) to (Q11);
				\draw [thick] (Q6) to (Q10);
				\draw [thick] (Q7) to (Q13);
				\draw [thick] (Q8) to (Q12);
				\draw [thick] (Q10) to (Q11);
				\draw [thick] (Q11) to (Q12);
				\draw [thick] (Q12) to (Q13);
				\draw [thick] (Q13) to (Q14);
				\draw [thick] (Q10) to (Q15);
				\draw [thick] (Q11) to (Q16);
				\draw [thick] (Q13) to (Q18);
				\draw [thick] (Q14) to (Q19);
				\draw [thick] (Q11) to (Q17);
				\draw [thick] (Q12) to (Q16);
				\draw [thick] (Q13) to (Q19);
				\draw [thick] (Q14) to (Q18);
				\draw [thick] (Q15) to (Q16);
				\draw [thick] (Q16) to (Q17);
				
			\end{tikzpicture}
		\end{adjustbox}
	}
	\subfigure[]{
	\begin{adjustbox}{width=0.4\textwidth}
		\begin{tikzpicture}[>=stealth]
			\node (Q0) at (0,6) [circle,draw=black!100,inner sep=2.2mm] {\Large0};
			\node (Q1) at (2,6) [circle,draw=black!100,inner sep=2.2mm] {\Large1};
			\node (Q2) at (4,6) [circle,draw=black!100,inner sep=2.2mm] {\Large2};
			\node (Q3) at (6,6) [circle,draw=black!100,inner sep=2.2mm] {\Large3};
			\node (Q4) at (8,6) [circle,draw=black!100,inner sep=2.2mm] {\Large4};
			\node (Q5) at (0,4) [circle,draw=black!100,inner sep=2.2mm] {\Large5};
			\node (Q6) at (2,4) [circle,draw=black!100,inner sep=2.2mm] {\Large6};
			\node (Q7) at (4,4) [circle,draw=black!100,inner sep=2.2mm] {\Large7};
			\node (Q8) at (6,4) [circle,draw=black!100,inner sep=2.2mm] {\Large8};
			\node (Q9) at (8,4) [circle,draw=black!100,inner sep=2.2mm] {\Large9};
			\node (Q10) at (0,2) [circle,draw=black!100,inner sep=1.8mm] {\Large10};
			\node (Q11) at (2,2) [circle,draw=black!100,inner sep=1.8mm] {\Large11};
			\node (Q12) at (4,2) [circle,draw=black!100,inner sep=1.8mm] {\Large12};
			\node (Q13) at (6,2) [circle,draw=black!100,inner sep=1.8mm] {\Large13};
			\node (Q14) at (8,2) [circle,draw=black!100,inner sep=1.8mm] {\Large14};
			\node (Q15) at (0,0) [circle,draw=black!100,inner sep=1.8mm] {\Large15};
			\node (Q16) at (2,0) [circle,draw=black!100,inner sep=1.8mm] {\Large16};
			\node (Q17) at (4,0) [circle,draw=black!100,inner sep=1.8mm] {\Large17};
			\node (Q18) at (6,0) [circle,draw=black!100,inner sep=1.8mm] {\Large18};
			\node (Q19) at (8,0) [circle,draw=black!100,inner sep=1.8mm] {\Large19};

			\draw [thick] (Q0) to (Q1);
			\draw [thick] (Q1) to (Q2);
			\draw [thick] (Q2) to (Q3);
			\draw [thick] (Q3) to (Q4);
			\draw [thick] (Q1) to (Q6);
			\draw [thick] (Q3) to (Q8);
			\draw [thick] (Q5) to (Q6);
			\draw [thick] (Q6) to (Q7);
			\draw [thick] (Q7) to (Q8);
			\draw [thick] (Q8) to (Q9);
			\draw [thick] (Q5) to (Q10);
			\draw [thick] (Q7) to (Q12);
			\draw [thick] (Q9) to (Q14);
			\draw [thick] (Q10) to (Q11);
			\draw [thick] (Q11) to (Q12);
			\draw [thick] (Q12) to (Q13);
			\draw [thick] (Q13) to (Q14);
			\draw [thick] (Q11) to (Q16);
			\draw [thick] (Q13) to (Q18);
			\draw [thick] (Q15) to (Q16);
			\draw [thick] (Q16) to (Q17);
			\draw [thick] (Q17) to (Q18);
			\draw [thick] (Q18) to (Q19);

		\end{tikzpicture}
	\end{adjustbox}
}
\subfigure[]{
		\begin{adjustbox}{width=0.8\textwidth}
		\begin{tikzpicture}[>=stealth]
			\node (Q0) at (-10,0) [circle,draw=black!100,inner sep=2.2mm] {\Large0};
			\node (Q1) at (-8,0) [circle,draw=black!100,inner sep=2.2mm] {\Large1};
			\node (Q2) at (-8,-2) [circle,draw=black!100,inner sep=2.2mm] {\Large2};
			\node (Q3) at (-8,-4) [circle,draw=black!100,inner sep=2.2mm] {\Large3};
			\node (Q4) at (-6,0) [circle,draw=black!100,inner sep=2.2mm] {\Large4};
			\node (Q5) at (-6,-4) [circle,draw=black!100,inner sep=2.2mm] {\Large5};
			\node (Q6) at (-4,2) [circle,draw=black!100,inner sep=2.2mm] {\Large6};
			\node (Q7) at (-4,0) [circle,draw=black!100,inner sep=2.2mm] {\Large7};
			\node (Q8) at (-4,-4) [circle,draw=black!100,inner sep=2.2mm] {\Large8};
			\node (Q9) at (-4,-6) [circle,draw=black!100,inner sep=2.2mm] {\Large9};
			\node (Q10) at (-2,0) [circle,draw=black!100,inner sep=1.8mm] {\Large10};
			\node (Q11) at (-2,-4) [circle,draw=black!100,inner sep=1.8mm] {\Large11};
			\node (Q12) at (0,0) [circle,draw=black!100,inner sep=1.8mm] {\Large12};
			\node (Q13) at (0,-2) [circle,draw=black!100,inner sep=1.8mm] {\Large13};
			\node (Q14) at (0,-4) [circle,draw=black!100,inner sep=1.8mm] {\Large14};
			\node (Q15) at (2,0) [circle,draw=black!100,inner sep=1.8mm] {\Large15};
			\node (Q16) at (2,-4) [circle,draw=black!100,inner sep=1.8mm] {\Large16};
			\node (Q17) at (4,2) [circle,draw=black!100,inner sep=1.8mm] {\Large17};
			\node (Q18) at (4,0) [circle,draw=black!100,inner sep=1.8mm] {\Large18};
			\node (Q19) at (4,-4) [circle,draw=black!100,inner sep=1.8mm] {\Large19};
			\node (Q20) at (4,-6) [circle,draw=black!100,inner sep=1.8mm] {\Large20};
			\node (Q21) at (6,0) [circle,draw=black!100,inner sep=1.8mm] {\Large21};
			\node (Q22) at (6,-4) [circle,draw=black!100,inner sep=1.8mm] {\Large22};
			\node (Q23) at (8,0) [circle,draw=black!100,inner sep=1.8mm] {\Large23};
			\node (Q24) at (8,-2) [circle,draw=black!100,inner sep=1.8mm] {\Large24};
			\node (Q25) at (8,-4) [circle,draw=black!100,inner sep=1.8mm] {\Large25};
			\node (Q26) at (10,-4) [circle,draw=black!100,inner sep=1.8mm] {\Large26};
			
			\draw [thick] (Q0) to (Q1);
			\draw [thick] (Q1) to (Q4);
			\draw [thick] (Q4) to (Q7);
			\draw [thick] (Q7) to (Q10);
			\draw [thick] (Q10) to (Q12);
			\draw [thick] (Q12) to (Q15);
			\draw [thick] (Q15) to (Q18);
			\draw [thick] (Q18) to (Q21);
			\draw [thick] (Q21) to (Q23);
			\draw [thick] (Q6) to (Q7);
			\draw [thick] (Q17) to (Q18);
			\draw [thick] (Q1) to (Q2);
			\draw [thick] (Q2) to (Q3);
			\draw [thick] (Q12) to (Q13);
			\draw [thick] (Q13) to (Q14);
			\draw [thick] (Q23) to (Q24);
			\draw [thick] (Q24) to (Q25);
			\draw [thick] (Q3) to (Q5);
			\draw [thick] (Q5) to (Q8);
			\draw [thick] (Q8) to (Q11);
			\draw [thick] (Q11) to (Q14);
			\draw [thick] (Q14) to (Q16);
			\draw [thick] (Q16) to (Q19);
			\draw [thick] (Q19) to (Q22);
			\draw [thick] (Q22) to (Q25);
			\draw [thick] (Q25) to (Q26);
			\draw [thick] (Q8) to (Q9);
			\draw [thick] (Q19) to (Q20);
			
		\end{tikzpicture}
	\end{adjustbox}
}
	
	\caption{(a)-(e) Manila, Jakarta, Tokyo, Almaden, and Kolkata of the IBMQ cloud platform, respectively.}
 \label{ibmqu}
\end{figure*}

\section{Parameter metrics}
\label{Parameter}
In this Appendix, we show the specific parameter metrics of the devices of each quantum computing cloud platform, such as T1, T2, error rate, etc. And each platform offers different parameters. It is important to note that these parameters change with calibration, so the specific values of these metrics are based on the time of the circuit demonstration. Tables~\ref{table4} to~\ref{tablesqp18} show the detailed parameters of the real devices, each table corresponding to an architecture. Table~\ref{tableibmq} provides only the average parameter metrics for the fake\_provider simulator on the IBMQ cloud platform.

	\begin{table}[]
		\centering
		\renewcommand{\arraystretch}{1.5} 
		\setlength{\tabcolsep}{1.1pt} 
		\caption{Calibration data for Fig.~\ref{fig2}(a), IBMQ\_quito.}
		\begin{ruledtabular}
		\begin{tabular}{@{}ccc@{}}
			Qubit1 & Qubit2 & CNOT error \\ \cline{0-2}
			Q0     & Q1     & 1.63E-02   \\
			Q1     & Q2     & 7.77E-03   \\
			Q1     & Q3     & 7.44E-03   \\
			Q3     & Q4     & 8.79E-03   \\ 
		\end{tabular}
	\end{ruledtabular}
     \label{table4}
	\end{table}

	\begin{table*}[]
		\centering
		\renewcommand{\arraystretch}{1.1} 
		\setlength{\tabcolsep}{1.5pt} 
		\caption{Calibration data for Wuyuan II of OriginQ cloud platform in China.}
		\begin{ruledtabular}
		\begin{tabular}{@{}ccccccc@{}}
			Qubit & T1($\mu s$)    & T2($\mu s$)   & Readout fidelity(F0/F1) & Single-qubit gate fidelity & Connection & CZ fidelity \\ \cline{0-6}
			Q0    & 12 & 2.7  & 0.971/0.897      & 0.9989                     & Q0\_Q1     & 0.9851      \\
			Q1    & 22 & 9.3  & 0.922/0.801      & 0.9989                     & Q1\_Q2     & 0.9619      \\
			Q2    & 4  & 7.1  & 0.978/0.883      & 0.9961                     & Q2\_Q3     & 0.7014      \\
			Q3    & 14 & 17.8 & 0.960/0.885      & 0.998                      & Q3\_Q4     & 0.8256      \\
			Q4    & 9  & 10.3 & 0.968/0.901      & 0.9987                     & Q4\_Q5     & 0.7132      \\
			Q5    & 16 & 3.3  & 0.944/0.868      & 0.9982                     & --         &   --         \\ 
		\end{tabular}
		\end{ruledtabular}
\label{WUyuan2}
	\end{table*}

\begin{table*}[]
		\centering
	\renewcommand{\arraystretch}{1.1} 
	\setlength{\tabcolsep}{1.5pt} 
	\caption{Calibration data for Wukong of OriginQ cloud platform in China.}
	\begin{ruledtabular}
	\begin{tabular}{@{}ccccccc@{}}
		Qubit &  T1($\mu s$)    & T2($\mu s$)   & Readout fidelity(F0/F1) & Single-qubit gate fidelity & Connection & CZ fidelity \\ \cline{0-6}
		Q40   & 26.007 & 0.532 & 0.8997           & 0.9973                     & Q40\_Q46   & 0.8813      \\
		Q45   & 26.58  & 0.354 & 0.8844           & 0.9905                     & Q45\_Q46   & 0.9744      \\
		Q46   & 23.758 & 0.707 & 0.8948           & 0.998                      & Q46\_Q52   & 0.97        \\
		Q48   & 13.582 & 0.339 & 0.8629           & 0.9963                     & Q48\_Q54   & 0.9637      \\
		Q52   & 19.289 & 0.977 & 0.929            & 0.9986                     & Q52\_Q53   & 0.9714      \\
		Q53   & 31.361 & 0.63  & 0.9454           & 0.9946                     & Q53\_Q54   & 0.978       \\
		Q54   & 22.699 & 5.941 & 0.863            & 0.9986                     & Q54\_Q60   & 0.9721      \\
		Q59   & 11.857 & 5.694 & 0.8859           & 0.9955                     & Q59\_Q60   & 0.9636      \\
		Q60   & 14.697 & 4.454 & 0.8798           & 0.9972                     & Q60\_Q66   & 0.9709      \\
		Q65   & 20     & 10    & 0.9175           & 0.9974                     & Q65\_Q71   & 0.9613      \\
		Q66   & 20     & 10    & 0.8695           & 0.9964                     &--         &  --            \\
		Q71   & 20     & 10    & 0.8806           & 0.9975                     & --        &   --           \\
	\end{tabular}
		\end{ruledtabular}
\label{WUkong}
\end{table*}

\begin{table*}[]
	\centering
	\renewcommand{\arraystretch}{1.1} 
	\setlength{\tabcolsep}{1.5pt} 
	\caption{Calibration data for ScQ-10 of Quafu cloud platform in China.}
	\begin{ruledtabular}
		\begin{tabular}{@{}cccccccc@{}}
			Qubit & T1($\mu s$)    & T2($\mu s$)    & Anharmonicity(GHZ) & Qubit frequency(GHZ) & Readout frequency(GHZ) & Connection & CZ fidelity     \\ \cline{0-7}
			Q1    & 26.79 & 2.33 & 0.25          & 5.31            & 6.6636            & Q2\_Q1     & 0.9787 \\
			Q2    & 29.38 & 1.82 & 0.207         & 4.681           & 6.6461            & Q3\_Q2     & 0.9564 \\
			Q3    & 58.96 & 2.31 & 0.246         & 5.367           & 6.6277            & Q4\_Q3     & 0.949  \\
			Q4    & 18.92 & 1.61 & 0.206         & 4.702           & 6.608             & Q5\_Q4     & 0.963  \\
			Q5    & 43.38 & 2.49 & 0.250682      & 5.299           & 6.5933            & Q6\_Q5     & 0.9669 \\
			Q6    & 47.37 & 2.14 & 0.203         & 4.531           & 6.5706            & Q7\_Q6     & 0.9663 \\
			Q7    & 37.88 & 1.97 & 0.2515        & 5.255           & 6.5541            & Q8\_Q7     & 0.956  \\
			Q8    & 19.5  & 1.33 & 0.2035        & 4.627           & 6.5318            & Q9\_Q8     & 0.9741 \\
			Q9    & 31.51 & 1.72 & 0.2462        & 5.275           & 6.5109            & Q9\_Q10    & 0.9909 \\
			Q10   & 23.43 & 3.56 & 0.208         & 4.687           & 6.4905            & --         & --      \\
		\end{tabular}
		\end{ruledtabular}
\label{tablesqp10}
	\end{table*}

\begin{table*}[]
	\centering
	\renewcommand{\arraystretch}{1.1} 
	\setlength{\tabcolsep}{1.1pt} 
	\caption{Calibration data for ScQ-18 of Quafu cloud platform in China.}
	\begin{ruledtabular}
	\begin{tabular}{@{}cccccccc@{}}
		Qubit & T1($\mu s$)    & T2($\mu s$)    & Anharmonicity(GHZ) & Qubit frequency(GHZ) & Readout frequency(GHZ) & Connection & CZ fidelity     \\ \cline{0-7}
		Q3    & 57.56365654 & 4.239166288 & 0.204783115   & 4.59            & 6.77627           & Q4\_Q3     & 0.989946019 \\
		Q4    & 27.88541771 & 3.699022109 & 0.191726174   & 5.02            & 6.75915           & Q5\_Q4     & 0.921217382 \\
		Q5    & 41.44235979 & 4.83465697  & 0.2026145     & 4.62            & 6.73673           & Q6\_Q5     & 0.903526443 \\
		Q6    & 40.10235639 & 2.298777179 & 0.198         & 5.071           & 6.71278           & Q7\_Q6     & 0.927608543 \\
		Q7    & 28.9943269  & 5.690175203 & 0.204836532   & 4.5             & 6.69334           & Q8\_Q7     & 0.976533411 \\
		Q8    & 23.60180849 & 2.203347967 & 0.19367078    & 4.982           & 6.675695          & Q9\_Q8     & 0.805179376 \\
		Q9    & 26.83896243 & 3.419803923 & 0.203920589   & 4.566           & 6.649799022       & Q9\_Q10    & 0.969400313 \\
		Q10   & 19.75997688 & 2.437594747 & 0.2           & 5.091           & 6.6334            & Q11\_Q10   & 0.9825224   \\
		Q11   & 23.21915934 & 3.533866245 & 0.206         & 4.646           & 6.625063867       & Q12\_Q11   & 0.965295259 \\
		Q12   & 27.30825285 & 1.324619262 & 0.2           & 5.038           & 6.647559341       & Q13\_Q12   & 0.973640043 \\
		Q13   & 41.15484398 & 2.77881436  & 0.204648089   & 4.605           & 6.665             & Q14\_Q13   & 0.95548674  \\
		Q14   & 28.11873582 & 1.493925899 & 0.198         & 4.993           & 6.69280357        & Q15\_Q14   & 0.9514352   \\
		Q15   & 45.11682637 & 2.953474165 & 0.206         & 4.545           & 6.705075651       & Q16\_Q15   & 0.971210129 \\
		Q16   & 29.79524557 & 1.568231794 & 0.198         & 4.869           & 6.725190179       & Q17\_Q16   & 0.966539068 \\
		Q17   & 45.11688811 & 5.466767893 & 0.204114512   & 4.578           & 6.751021438       & Q18\_Q17   & 0.982567594 \\
		Q18   & 37.2792584  & 3.068254432 & 0.196         & 5.048           & 6.770705353       & Q19\_Q18   & 0.979586641 \\
		Q19   & 40.83724158 & 5.153026625 & 0.203098846   & 4.682           & 6.787665          & Q20\_Q19   & 0.984410078 \\
		Q20   & 41.42844931 & 3.116332126 & 0.196583622   & 5.125           & 6.807017363       &     --        &   --           
	\end{tabular}
		\end{ruledtabular}
\label{tablesqp18}
\end{table*}

\begin{table*}[]
	\centering
	\renewcommand{\arraystretch}{1.1} 
	\setlength{\tabcolsep}{1.1pt} 
	\caption{Calibration data for architectures on IBMQ cloud platform.}
	\begin{ruledtabular}
	\begin{tabular}{@{}ccccc@{}}
		Architecture & T1($\mu s$)    & T2($\mu s$)      & Single-qubit gate error & CNOT error \\ \cline{0-4}
		Manila       & 0.1125 & 0.0372 & 0.0011                     & 0.0116        \\
		Quito        & 0.0615 & 0.0518 & 0.0017                     & 0.0354        \\
		Jakarta      & 0.1284 & 0.0339 & 0.0003                     & 0.0126        \\
		Guadalupe    & 0.0702 & 0.0881 & 0.0004                     & 0.0108        \\
		Tokyo        & 0.0876 & 0.0519 & --                         & 0.0313        \\
		Almaden      & 0.0868 & 0.0643 & --                         & 0.0238        \\
		Kolkata      & 0.1099 & 0.0968 & 0.0003                     & 0.0109        \\ 
	\end{tabular}
	\end{ruledtabular}
\label{tableibmq}
\end{table*}


	\end{document}